\documentclass[a4paper,fleqn,usenatbib]{mnras}

\usepackage{newtxtext,newtxmath}

\usepackage[T1]{fontenc}
\usepackage{ae,aecompl}

\usepackage{graphicx}	
\usepackage{amsmath}	
\usepackage{amssymb}	
\usepackage{threeparttable}
\usepackage{bm}
\usepackage{booktabs}

\title{Intraday Optical Variability of BL Lacertae}

\author[Meng et al.]{Nankun Meng,$^{1}$
Jianghua Wu,$^{1}$
\thanks{E-mail: jhwu@bnu.edu.cn}
James R. Webb,$^{2}$
Xiaoyuan Zhang,$^{1}$
Yan Dai$^{1}$
\\
$^{1}$Department of Astronomy, Beijing Normal University, 100875, Beijing, China\\
$^{2}$Department of Physics, Florida International University, Miami, FL 33199, USA
}

\date{Accepted XXX. Received YYY; in original form ZZZ}

\pubyear{2016}

\begin{document}
\label{firstpage}
\pagerange{\pageref{firstpage}--\pageref{lastpage}}
\maketitle

\begin{abstract}

We monitored BL Lacertae simultaneously in the optical $B, V, R$ and $I$ bands for 13 nights during the period 2012-2016. The variations were well correlated in all bands and the source showed significant intraday variability (IDV). We also studied its optical flux and colour behaviour, and searched for inter-band time lags. A strong bluer-when-brighter chromatism was found on the intra-night time-scale. The spectral changes are not sensitive to the host galaxy contribution. Cross-correlation analysis revealed possible time delay of about 10 min between variations in the $V$ and $R$ bands. We interpreted the observed flares in terms of the model consisting of individual synchrotron pulses.

\end{abstract}




\section{Introduction}

BL Lacertae is the archetype of BL Lac class, which together with flat spectrum radio quasars (FSRQs), constitutes a violently variable class of active galactic nuclei (AGN) known as blazars.
Blazars are characterized by high and variable polarization, synchrotron emission from relativistic jets, core-dominated radio morphology and intense flux and spectral variability in all wavelengths ranging from radio to
 $\gamma$-ray on a wide variety of time-scales \citep{1995PASP..107..803U,1995ARA&A..33..163W,2003ApJ...596..847B}. 
 Blazar variations can be divided into long-term variability, short-term variability and microvariability. Magnitude changes of few hundredth to tenths over a day are called intraday variability (IDV) or microvariability \citep{1989Natur.337..627M,2004AJ....127...17H,2015MNRAS.450..541A}.
 
 BL Lacertae, located at redshift value of z = 0.0668$\pm$0.0002 \citep{1977ApJ...212L..47M}, is hosted by a giant elliptical galaxy with R=15.5 \citep{2000ApJ...532..740S}. As its first spectral component peaks in near-IR (NIR)/optical region, BL Lacertae is a typical low-frequency peaked blazar (LBL) with the radio to X-ray spectral index equal to 0.84 \citep{1998MNRAS.299..433F,2004A&A...419...25F}.
 
BL Lacertae was continuously observed by several multiwavelength campaigns carried out by the Whole Earth Blazar Telescope (WEBT/GASP) \citep{2003ASPC..299..221V,2004A&A...424..497V,2006A&A...456..105B,2009A&A...507..769R}. Numerous investigations have been carried out to search for the flux variations, spectral changes and periodicities \citep{1972ApJ...178L..51E,1992AJ....104...15C,2002A&A...390..407V,2003A&A...397..565P,2015MNRAS.450..541A,2015MNRAS.452.4263G}. The majority of the observations revealed that its IDV amplitude is larger at higher frequencies and decreased as the flux increased. These studies confirmed the presence of bluer-when-brighter (BWB) trend, but yielded no evidence for periodicities \citep{1992AJ....104...15C,2002A&A...390..407V,2015MNRAS.452.4263G}. \citet{2002AJ....124.3031H} proposed the evidence for periodicity of 308 d for total flux variations in 22 yr. Most authors did not find any significant time lags between different optical bands during a single night observation \citep{1998A&A...332L...1N}. However, \citet{2003A&A...397..565P} found that the delay between $B$- and $I$- band light curves was $\sim$ 0.4 h. In addition, a possible time-lag between $e$ and $m$ bands was reported as $\sim$ 11.6 min by \citet{2006MNRAS.373..209H}. 

Polarimetric monitoring can offer information about the physical processes in blazar. Unlike radio polarization, optical polarimetry probes the central nuclear regions of blazar jets \citep{2014A&ARv..22...73F}. The optical polarized emission, as a potential good tracer of the high-energy emission, is being widely studied \citep{2008Natur.452..966M,2013MNRAS.436.1530R,2013ApJS..206...11S,2015A&A...578A..68C}. Since blazar optical radiation is dominated by the synchrotron mechanism, this implies the presence of highly ordered large scale magnetic fields \citep{1959ApJ...130..241W}. Observation of optical polarizations of BL Lac objects has shown the degree of polarization varies on time-scales from IDV to years \citep{2000AJ....119.1542I,2001A&A...376...51T,2008PASJ...60L..37S}. 
Occasionally, the correlation \citep{2008ApJ...672...40H} and anticorrelation \citep{2014ApJ...781L...4G} between the total and polarized fluxes is observed, while in general no clear relation is found.

In this paper, we studied the IDV and colour behaviour of this object, and searched for the inter-band time lags with data collected during the period $2012-2016$ with high temporal resolution. The results can give us insight into theoretical causes of variability. This paper is organized as follows: Section 2 describes the observations and data reductions, and Section 3 provides a brief introduction to various analysis techniques, followed by the results in Section 4. The discussion and conclusions are in Sections 5 and 6, respectively.

\begin{figure}
\label{fig:standstars} 
	\includegraphics[scale=0.6]{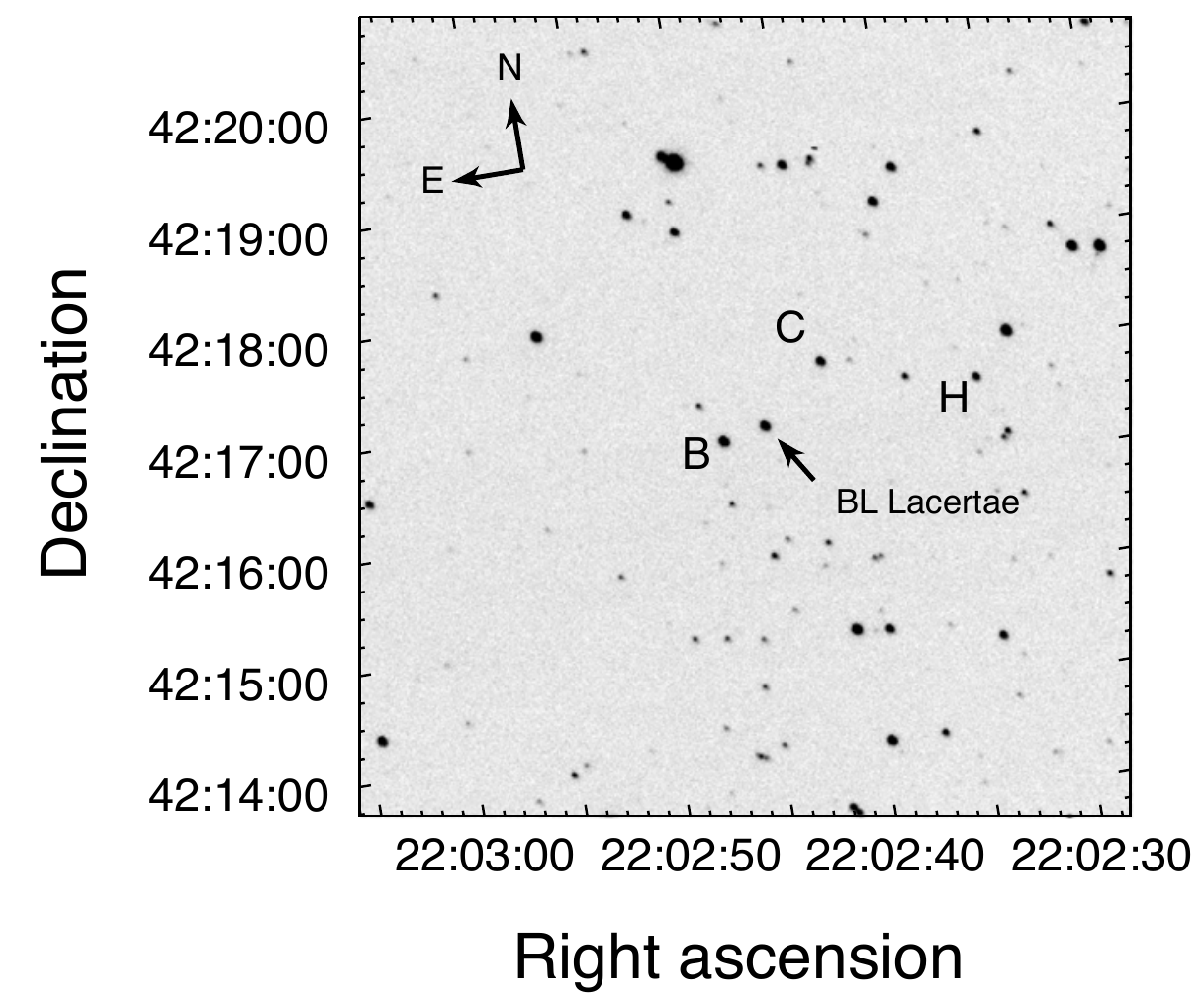}
    \caption{Finding chart of BL Lacertae in the $B$ band on 2015 October 19.}
\end{figure}

\begin{figure*}
	\includegraphics[scale=0.49]{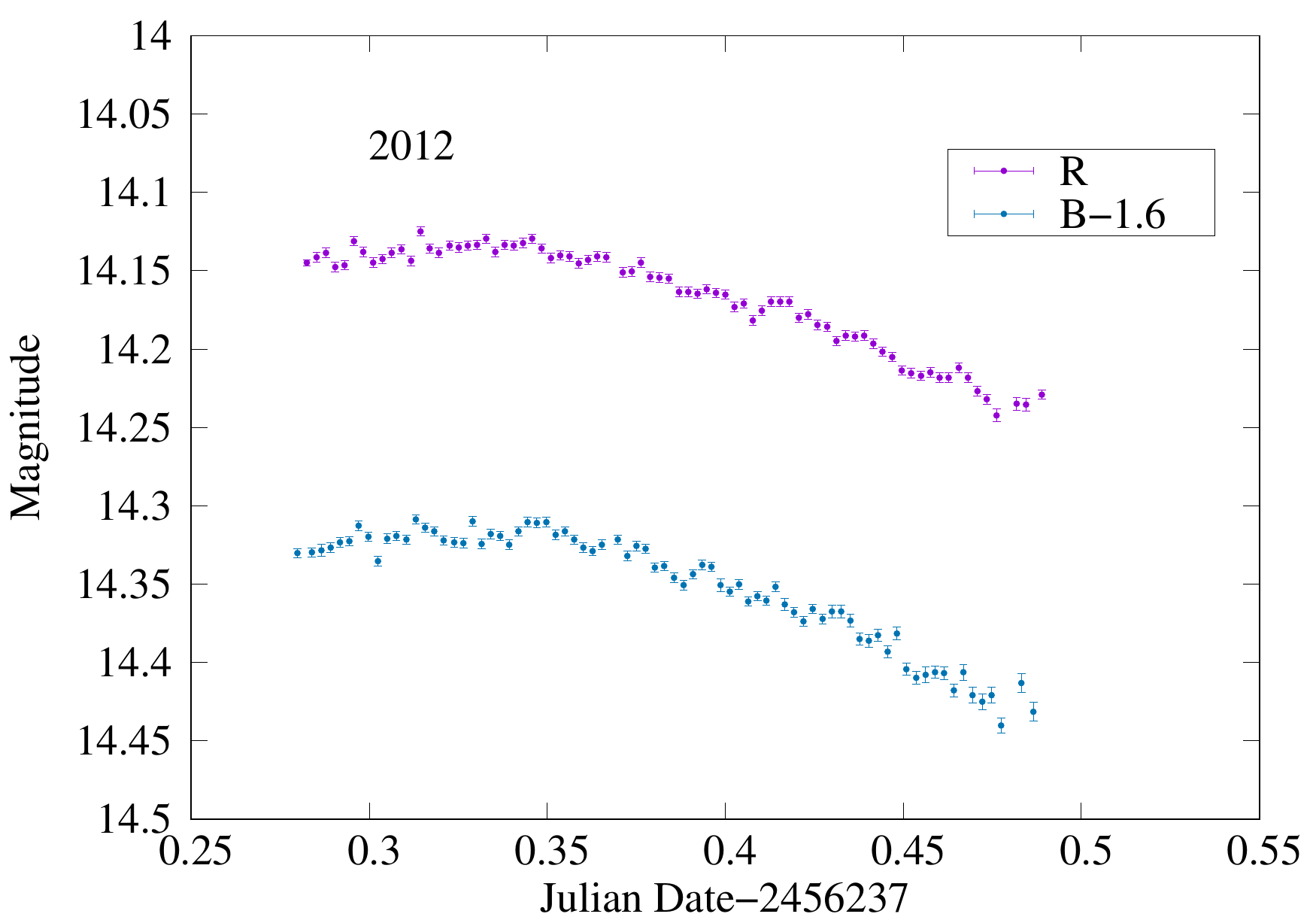}
	\includegraphics[scale=0.49]{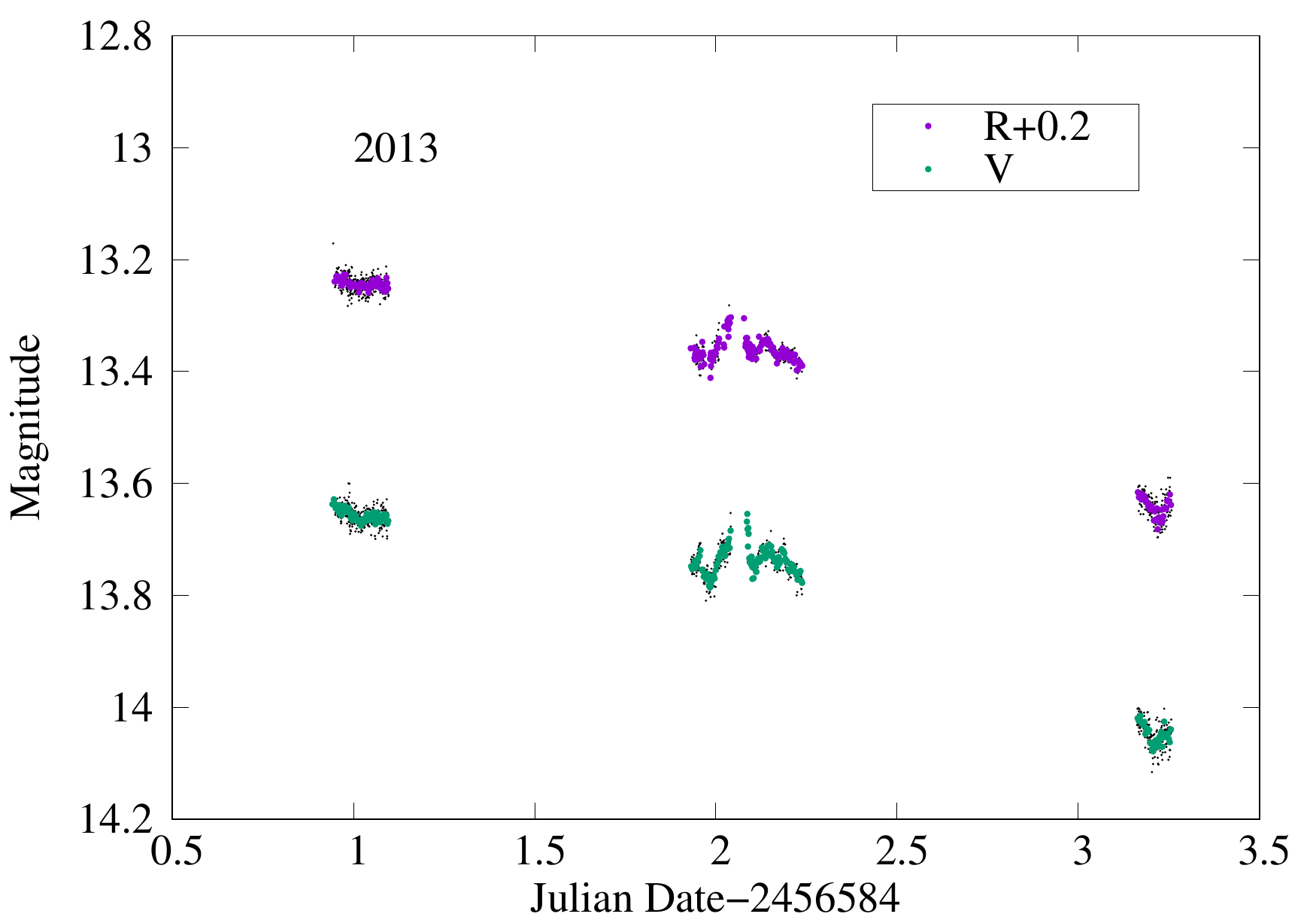}
	\includegraphics[scale=0.49]{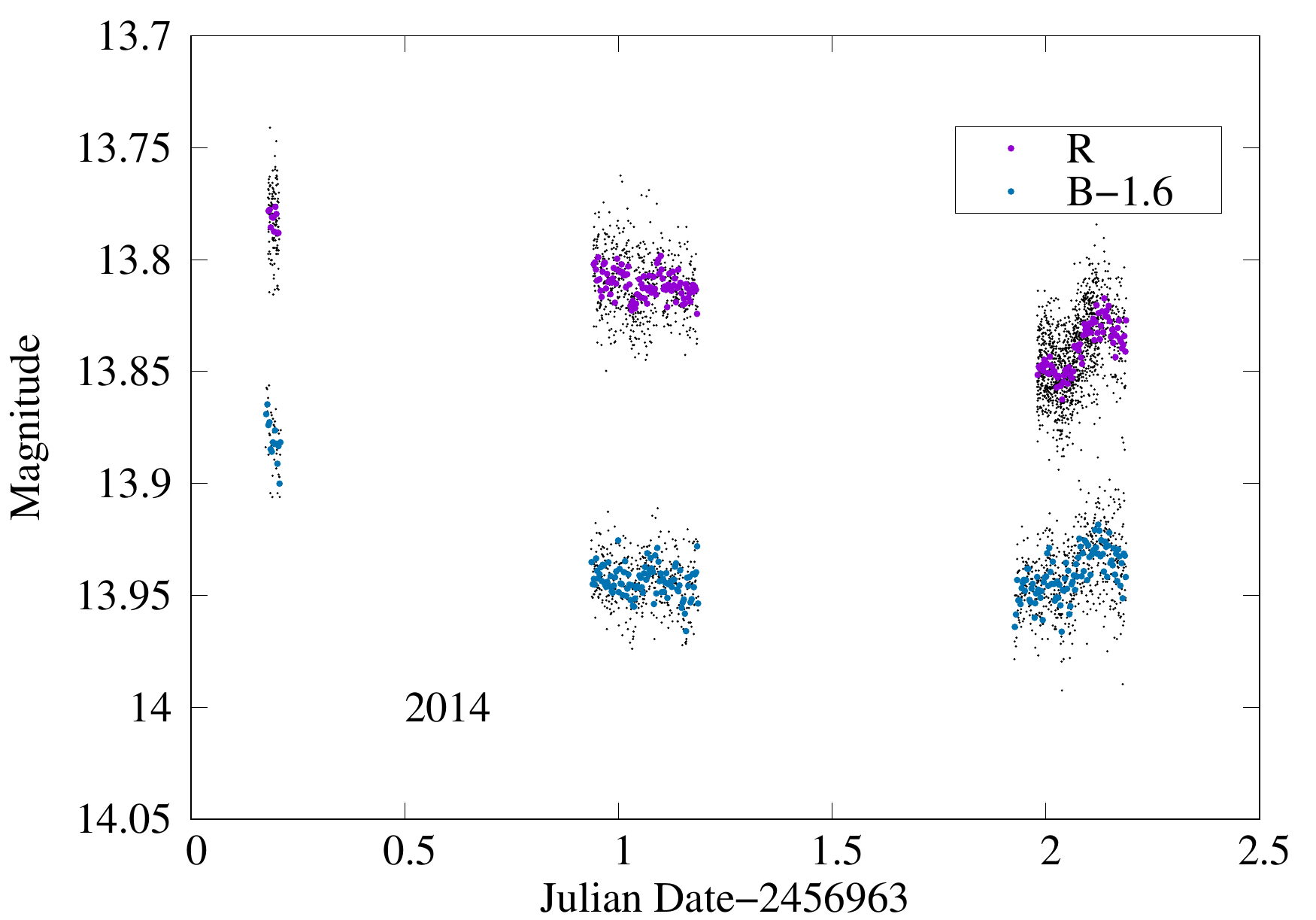}
	\includegraphics[scale=0.49]{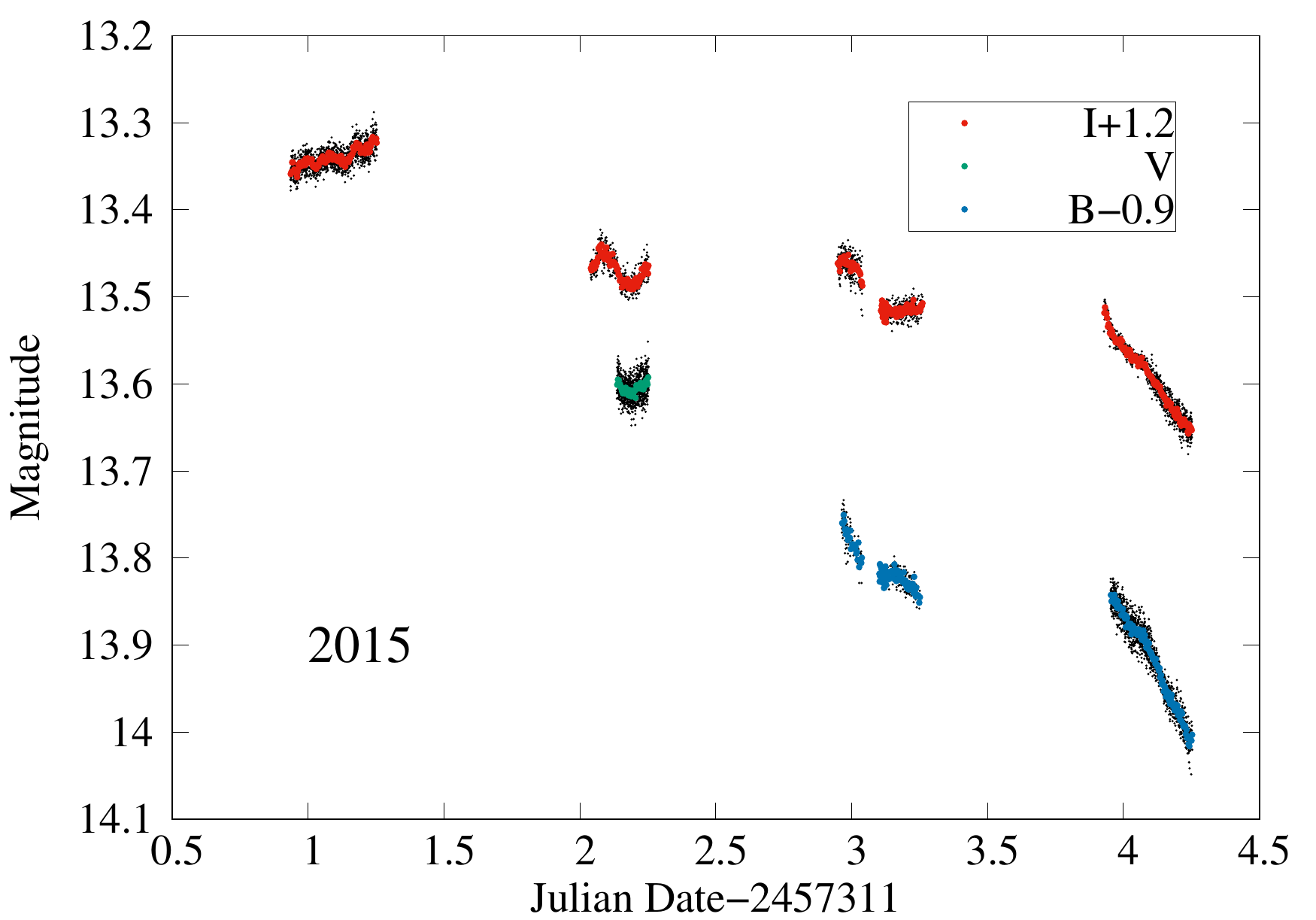}
	\includegraphics[scale=0.52]{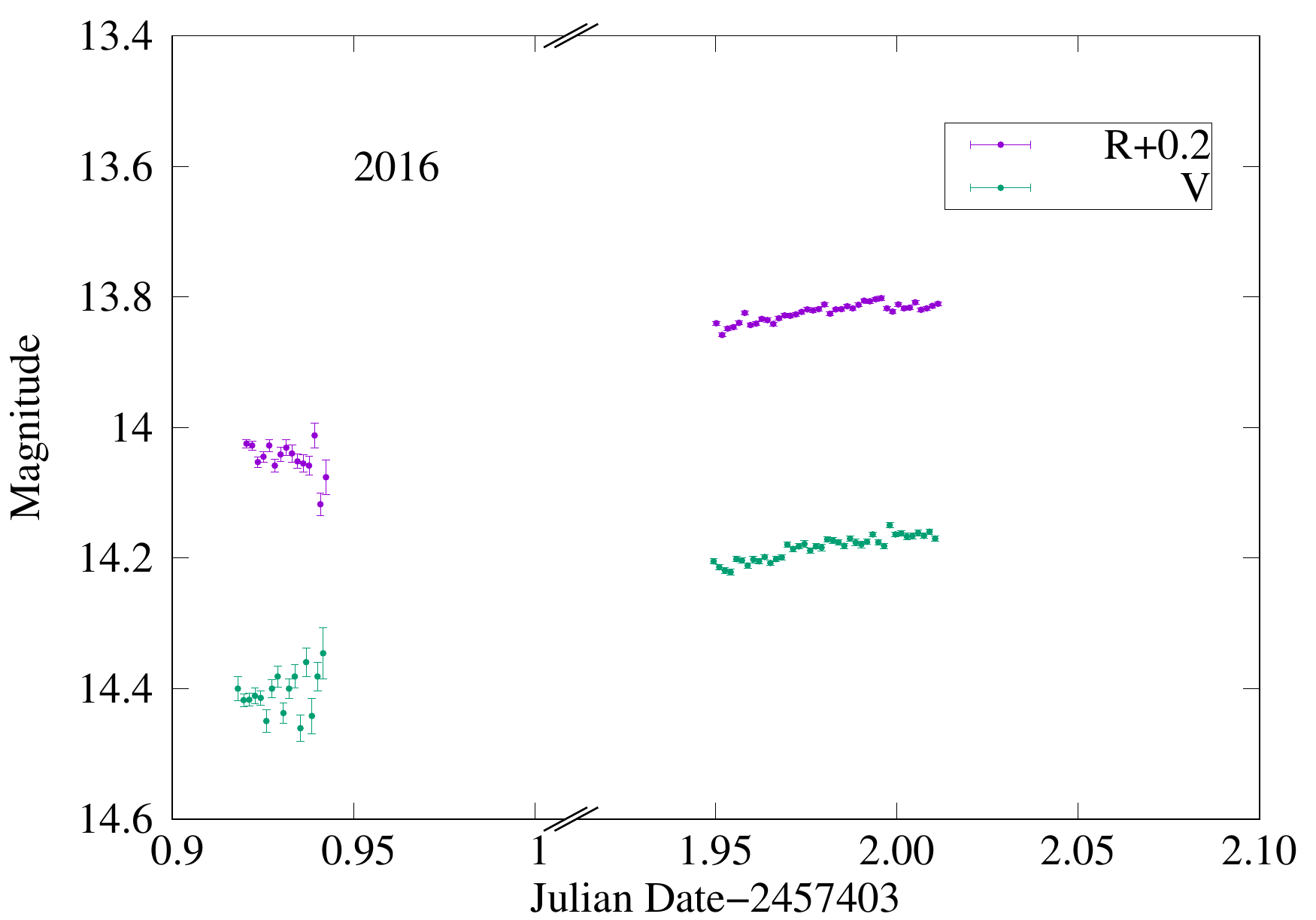}	
	
    \caption{Light curves of BL Lacertae in the $B, V, R$ and $I$ bands in 5 yr. Small black dots denote original data. Different colour dots denote smoothed data in different passbands. For clarity, several band light curves are shifted correspondingly.}
    \label{fig:lightcurves}
\end{figure*}	

\section{OBSERVATIONS AND DATA REDUCTIONS}

The monitoring was performed with three telescopes at Xinglong Station of the National Astronomical Observatories Chinese Academy of Sciences (NAOC). The details of the telescopes are shown in Table~\ref{tab:telescopes}. We observed BL Lacertae in the $B, V, R$ and $I$ bands for 13 nights covering the period from 2012 November 5 to 2016 January 17. During 12 of those nights, we observed simultaneously in different bands, providing a total of 24 intraday light curves. The longest individual duration of observation was about 7.7 h. The entire observation log together with all the available results is presented in Table~\ref{tab:results}. More than ten thousand original data points were collected on the 13 nights.

The data reduction procedures included bias subtraction, flat-fielding, extraction of instrumental magnitudes and flux calibration. The pre-processing of the raw data was accomplished by using standard procedures in the IRAF \footnote{IRAF is distributed by the National Optical Astronomy Observatories, which are operated by the Association of Universities for Research in Astronomy, Inc., under cooperative agreement with the National Science Foundation.} software. For each night, photometry was carried out  with five different aperture radii, i.e., $\sim$ $1\times$ FWHM,  $1.5\times $FWHM, $2\times$ FWHM, $3\times$FWHM, $4\times$ FWHM. The minimum standard deviation of photometric error corresponds to the best aperture and we finally selected the best aperture data for our analysis. Three local comparison stars (B, C, H in Fig.~\ref{fig:standstars}) were observed in the same field. The standard magnitudes of these stars in the $B, V, R$ and $I$ bands are given by \citet{1985AJ.....90.1184S}. The brightness of BL Lacertae was calibrated relative to the average brightness of stars B and C. They have similar magnitude and colour to BL Lacertae. Star H acted as a check star. 

The host galaxy of BL Lacertae is relatively bright and its contribution to the magnitudes was subtracted after flux calibration in order to avoid contamination. \citet{2007A&A...476..723H}  derived a $B$ magnitude of 17.35 for the host galaxy of BL Lacertae. The host galaxy contribution in $V, R$ and $I$  bands was inferred by adopting the elliptical galaxy colours of $B-V=0.99$, $V-R=0.59$ and $V-I=1.22$ from  \citet{2001MNRAS.326..745M}.  \citet{2002A&A...390..407V} estimates that the host galaxy contribution to the observed flux is about 60\% of  the whole galaxy flux. The magnitudes were transformed into flux and the host galaxy contribution was removed.

Given that short exposure times cause the data dispersion, a smoothing algorithm was implemented for the data in 2013, 2014 and 2015. The data were smoothed using the average of each 3 mins bin. The overall light curves of this source are displayed in Fig.~\ref{fig:lightcurves}. Small black dots denote original data and different colour dots denote smoothed data in different passbands. There are about 2000 smoothed data points used for further calculation. For clarity, several band light curves are shifted correspondingly.

\begin{table}
	\centering
	\caption{Parameters of three telescopes.}
	\label{tab:telescopes} 
	\begin{tabular}{lcr} 
		\hline
		Telescope & CCD resolution & CCD view\\
		\hline
		60cm reflector & 1".06 /pixel & $11'\times11'$\\
	        85cm reflector & 1" /pixel &  $33'\times33'$ \\
                216cm reflector    & 0".305 /pixel &  $6'.5\times5'.8$\\
		\hline
	\end{tabular}
\end{table}


\begin{table*}
	\centering
	\caption{Results of BL Lacertae.}
	\label{tab:results} 
        \begin{threeparttable}  
	
	\begin{tabular}{lccccccccccc} 
		\hline
		Date (yyyy mm dd) & Telescope & Band & No. of data points & Durations (h) &  \multicolumn{3}{c}{$\chi ^2$ test}  & \multicolumn{3}{c}{ANOVA test}  & A (\%) \\
		\hline
		 & & & & & $F$ & CV  & Var? & $F$ & CV & Var?  &    \\
		\hline
		2012 11 05  & 216cm                  & $B$       &  77   & 4.960  & 2884  &  107.6 &  Y    & 26.26  & 3.012 & Y    &    12.12  \\
		                    &                             & $R$       &  77   & 4.954  & 3821  & 107.6 & Y     &  37.96  & 3.012 & Y      &     10.77 \\
		2013 10 19  & 85cm                   & $V $      &  60    &  3.698  & 76.65 & 87.17  & N    &   2.471 & 3.536 & N     &               \\
		                   &                              & $ R$     &  59   &  3.946   & 502.3  & 85.95 & Y     &  0.674 & 3.791 & N       &                  \\   
		2013 10 20 & 85cm                    & $V $      &  119  &  7.354  &  509.4 & 156.7 & Y     & 3.268  & 2.358 & Y      &      12.73 \\
		                   &                              & $ R$      &  119  & 7.397  &  453.4 & 156.7 & Y     &  2.707 & 2.358 & Y     &    10.42   \\
		2013 10 21 & 85cm                    &  $V$      & 36    & 2.199   &  40.41 & 57.34 & N     & 2.645 & 5.944 & N       &               \\
		                   &                              &  $R$     & 36   & 2.186    & 37.59  & 57.34 & N       & 3.022  & 5.944 & N     &               \\
	        2014 11 01 & 60cm 85cm           & $B $      & 14     & 0.802  & 4.407 & 27.69 & N       & 1.103 & 99.40 & N       &               \\
	                           &                              & $R $      & 10     &  0.559  &  0.830 & 21.66 & N       & 0.253 & 99.40 & N       &               \\
	        2014 11 02 & 60cm 85cm           & $B $      &  97    & 5.981  &   11.54 & 131.1 & N      & 0.428 & 2.608 & N       &               \\
	                          &                                &$ R$      & 94    & 5.798  &  17.97 & 127.6 & N      &  0.684 &  2.690 & N       &                \\
	        2014 11 03 & 60cm 85cm            &$ B  $    & 101  & 6.230  &   22.45 & 135.8 & N      & 1.517  & 2.535 & N       &                \\
	                           &                                &$ R $    & 81    & 4.968  & 32.67 & 112.3 & N       &   5.749 & 2.889 & Y     &            \\
	        2015 10 16 & 60cm                     & $I    $  &123     & 7.589 &  40.93 & 161.3 & N      & 6.820 &  2.310 & Y     &            \\
	        2015 10 17 &60cm 85cm             &$ V  $   & 45    & 2.742   &  6.795 & 68.71 & N      &  1.889 & 4.539 & N         &           \\
	                           &                               &$ I $      & 83        & 5.094 &  50.11 & 114.7 & N       & 10.58 & 2.889 & Y      &         \\
	        $2015\ 10\ 18_1$ & 60cm 85cm &$ B$      & 29    & 1.740  & 45.90 & 48.28 & N         &  4.102 & 9.449 & N       &         \\
	                           &                               &$ I $      & 36      & 2.162 & 9.297 & 57.34 & N         & 1.846 & 5.944 & N       &       \\
	         $2015\ 10\ 18_2$ & 60cm 85cm &$ B$     &  82    & 3.568   &  91.27 & 113.5 & N        &  1.400 & 2.889 & N        &       \\
	                           &                                &$ I $      &  85  & 3.665    & 14.65 & 117.1 & N        &  0.320 & 2.783 & N         &         \\                   
	        2015 10 19 & 60cm 85cm           & $B $      & 116   & 7.171  & 1603 & 153.2 & Y      &  134.6  & 2.358 & Y         &      15.85 \\
	                           &                               & $I  $      & 125  & 7.722  &  771.4 &  163.6 & Y      & 135.0  & 2.266 & Y        &     13.49  \\
	        2016 01 16 & 85cm                    & $V  $     & 16    & 0.565   & 15.11 & 30.58 & N        & 0.023 & 26.87 & N         &                \\
	                           &                              & $R  $     & 15    & 0.528   & 21.03 & 29.14 & N        &  0.365 & 26.87 & N       &                \\
	        2016 01 17 & 85cm                    & $V  $     & 40  & 1.468     & 319.5 & 62.43 & Y       & 5.732 & 5.116 & Y     &     6.72    \\
	                           &                             & $R   $    & 40    & 1.469    &  349.6 & 62.43 & Y        & 6.079 & 5.116 & Y     &     5.31    \\
		\hline
	\end{tabular}
\begin{tablenotes}
	\item[Note.] We displayed two separate results ($2015\ 10\ 18_1$ and  $2015\ 10\ 18_2$) because of the existence of gaps in the light curves on 2015 October 18.
	\end{tablenotes}
\end{threeparttable}
\end{table*}

\begin{table*}
	\centering
	\caption{colour variations fitting results.}
	\label{tab:colourfits} 
	\begin{threeparttable}
	
	\begin{tabular}{llrrrrrrrrrr} 
		\hline
	         Date ( yyyy mm dd ) & CI & $\rm <CI>$ & $\rm <SI>$ & Slope & Intercept & r & Sc & Probability & F & $\rm F_{cri}$ & Plot  \\
	         (1)&(2)&(3)&(4)&(5)&(6)&(7)&(8)&(9)&(10)&(11)&(12)\\
		\hline
		2012 11 05 & $B-R$    & 1.785 & 4.387 & 0.048  &  1.767  &  0.298 &  0.475 &  0.010 &  8.081 & 4.351 & Yes\\
		2014 11 01 & $B-R$    & 1.699 & 4.177 & 0.133  &  1.674  &  0.103 &  0.428 &  0.482 &  0.577 &  5.987 & No\\
		2014 11 02 & $B-R$    & 1.732 & 4.258 & $-$0.005  &  1.738  &  0.003 & $-$0.084 &  0.689 &  0.162 &  3.996& No\\
		2014 11 03 & $B-R$   & 1.699 &  4.176 & 0.048  &  1.647  &  0.343 &  0.567 &  0.005 &  9.923 &  4.351 & Yes\\
		2015 10 $18_1$ & $B-I$    & 2.416 & 3.723 & 0.313  &  2.104  &  0.478 &  0.674 &  0.002 &  13.72 &  4.494 & Yes\\
		2015 10 $18_2$ & $B-I$    &  2.410 & 3.713 & 0.137  &  2.251  &  0.390 &  0.456 &  0.000 &  26.88 &  4.067 & Yes\\
		2015 10 19 & $B-I$    & 2.422 & 3.732 & 0.179  &  2.225  &  0.917 &  0.956 &  0.000 &  343.3 &  4.149 & Yes\\
		2013 10 19 & $V-R$    & 0.614 & 3.729 & 0.049  &  0.563  &  0.077 &  0.207 &  0.236 &  1.507 &  4.381 & No\\
		2013 10 20 & $V-R$    & 0.574 & 3.486 & $-0.039$  &  0.617  &  0.028 &  $-0.214$ &  0.163 &  1.993 &  3.976 & No\\
		2013 10 21 & $V-R$    &  0.607 & 3.690 & 0.003  &  0.607  &  0.000 &  0.000 &  0.979 &  0.001 &  4.301& No\\
		2016 01 16 & $V-R$    & 0.539 & 3.280 & $-7.111$  &  7.159  &  0.886 &  $-0.886$ &  0.005 &  30.93 &  6.608 & Yes \\
		2016 01 17 & $V-R$    & 0.562 & 3.416 & $-0.561$  &  1.112  &  0.479 &  $-0.701$ &  0.002 &  13.84 &  4.494 & Yes\\
		2015 10 17 & $V-I$    & 1.326 & 3.259 & 0.078  &  1.311  &  0.233 &  0.329 &  0.112 &  3.044 &  4.844 & No\\
		\hline
	\end{tabular}
	\begin{tablenotes}
	\item[\textbf{Notes.}] (1) Date; (2) colour; (3) the average value of different colours (in mag); (4) the average optical spectral index; (5) colour-time correlation slope; (6) colour-time correlation intercept; (7) correlation coefficient; (8) Spearman correlation coefficient; (9) null hypothesis probability; (10) $F$-test results; (11) critical value of $F$-test; (12) if the results are greater than the critical value of $F$-test, we will plot them in Fig~\ref{fig:colourtime}. 
	\end{tablenotes}
\end{threeparttable}
\end{table*}

\section{Variability Detection Criteria}

To quantify the IDV of BL Lacertae, two statistical analysis techniques were adopted, the $\chi ^2$ test and ANOVA test.

\subsection{$\chi ^2$ test}

The $\chi ^2$ statistic is defined as:
\begin{equation}
    \chi ^2 = \sum_{i=1}^N \frac{(V_i - \overline{V})^2}{ \sigma^2_i},
\end{equation}
where, $\overline{V}$ is the mean magnitude of all $i$th observation $V_i$ with a corresponding error $\sigma_i$. The exact errors from the IRAF reduction package are smaller than the real ones by a factor of 1.3 to 1.75 \citep{2008AJ....135.1384G,2015MNRAS.450..541A}. We chose the factor of 1.5 for data processing to get a better estimate of the actual photometric errors. If the actual variability is greater than the critical value at the $N-1$ degree of freedom and selected significance level, then the presence of variability can be claimed.

\subsection{ANOVA test}

\citet{1998ApJ...501...69D} used the one-way ANOVA to investigate the variability of quasars. The mathematical description of the one-way ANOVA test is as followed: if $y_{ij}$ represents the $i$th (with $ i=1, 2 \ldots ,n_j$) observation on the $j$th (with $j=1, 2,\ldots ,k$) group, the linear model describing every observation is
\begin{equation}
    y_{ij} = \overline{y}+g_i+\varepsilon_{ij},
\end{equation}
where, $\overline{y}$ represents the mean value of the whole data set, $g_j=\overline{y}_j -\overline{y}$ the between-groups deviation and $\varepsilon_{ij}=y_{ij} - y_i$ the within-groups deviation. Our observations on each night were divided into groups of five consecutive data points. The total sample variation can be separated between and within group deviations:

\begin{equation}
    \sum_{j=1}^{k} \sum_{i=1}^{n_j} (y_{ij} -\overline{y})^2 = \sum_{j=1}^{k} (y_i -\overline{y})^2 + \sum_{j=1}^{k} \sum_{i=1}^{n_j} (y_{ij} -\overline{y_j})^2.
	\label{eq:anova}
\end{equation}
Equation~(\ref{eq:anova}) can be shortened to $SS_{\rm T}=SS_{\rm G}+SS_{\rm R}$. $SS_{\rm T}$ stands for the total sum of squares that describes the total deviation of the data with respect to the mean. $SS_{\rm G}$ stands for the right-hand side of equation~(\ref{eq:anova}) and $SS_{\rm R}$ is the total error.

The statistics corresponds to the $F$ distribution with $k-1$ and $N-k$ degrees of freedom.
\begin{equation}
   F= \frac{SS_{\rm G}/(k-1)}{SS_{\rm R}/(N-k)}.
\end{equation}
For a certain significance level, if $F$ value exceeds the critical value, the null hypothesis will be rejected and implying the existence of variability.

\begin{figure}
	\centering
	\includegraphics[scale=0.23]{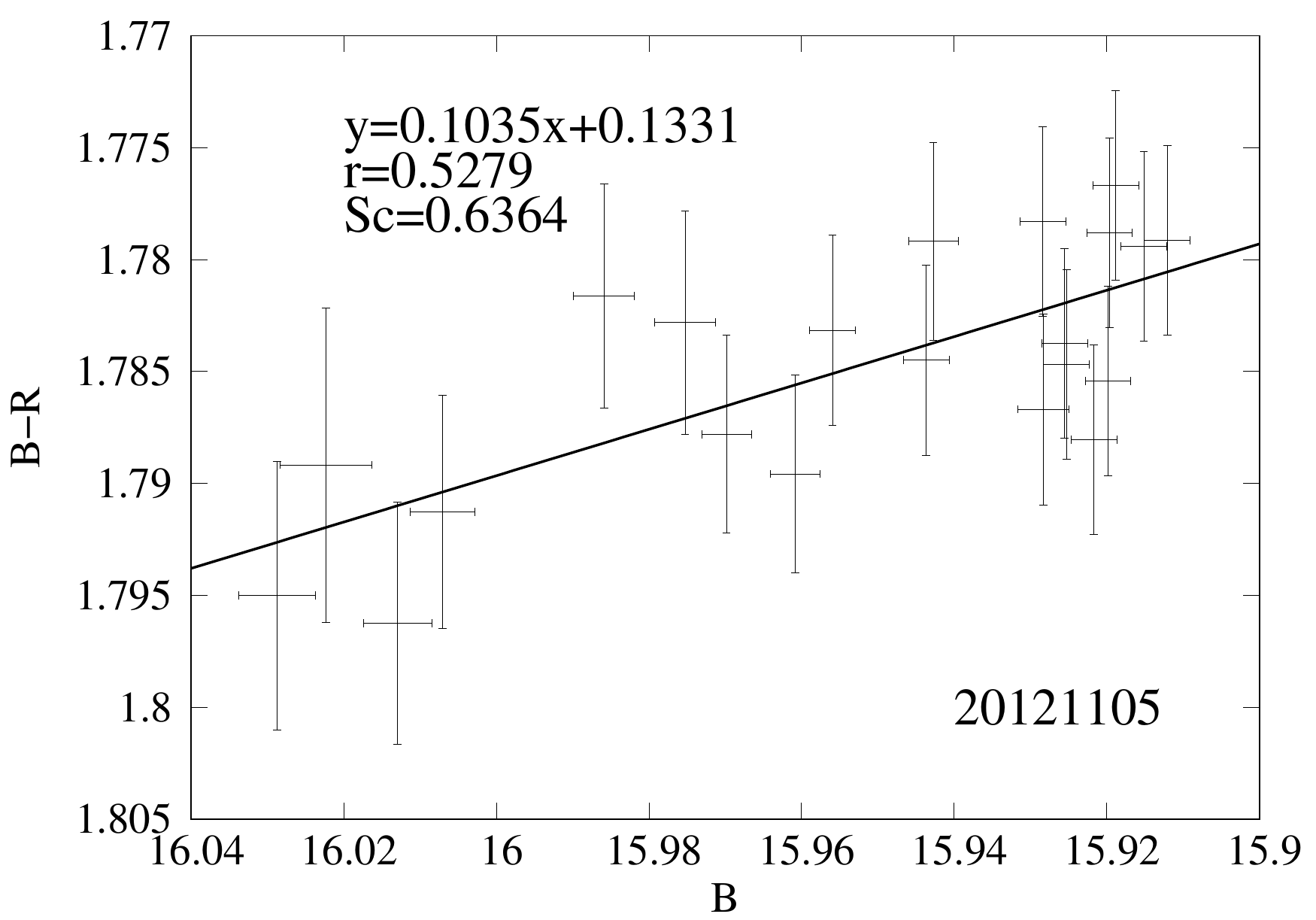}
	\includegraphics[scale=0.23]{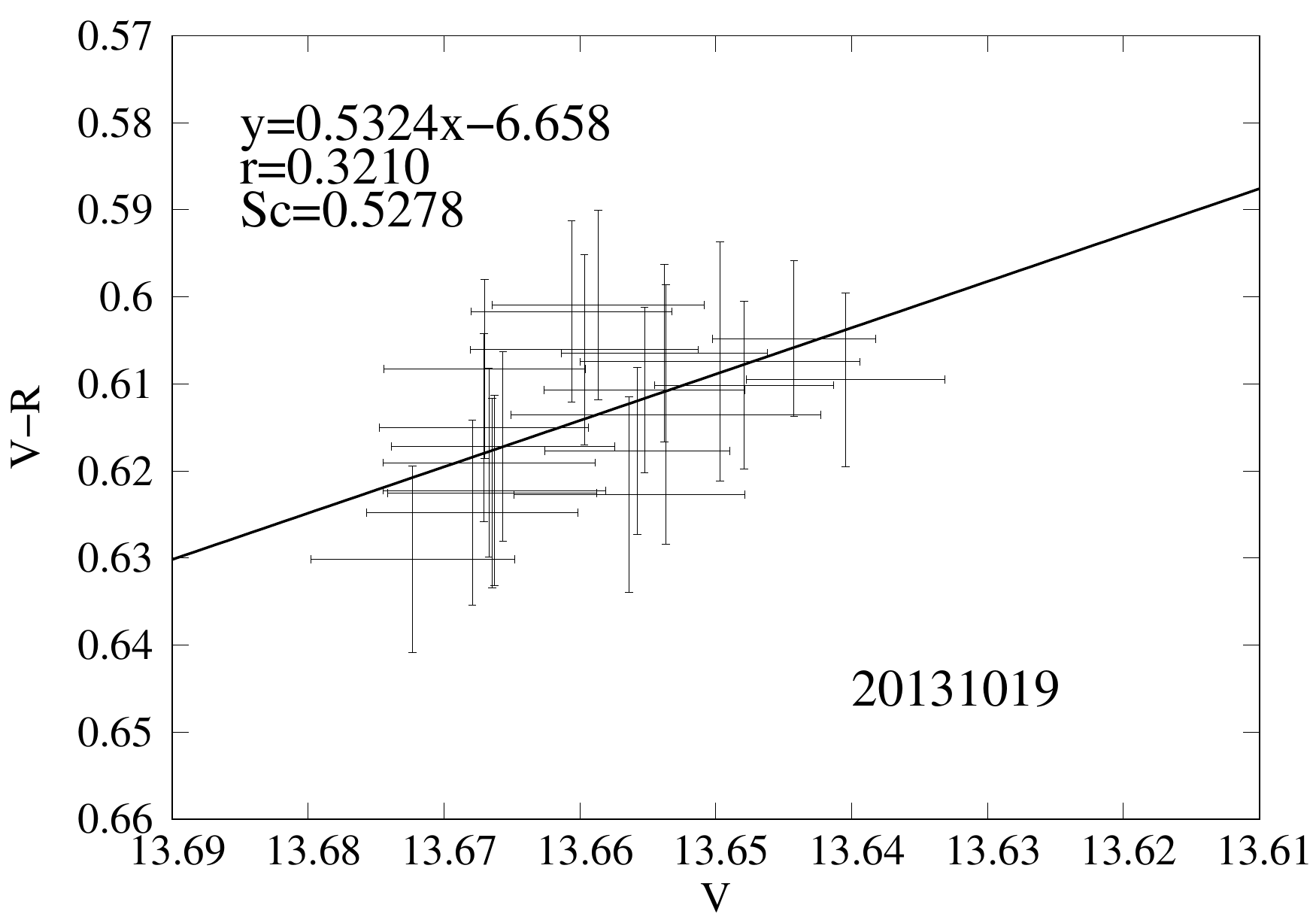}
	\includegraphics[scale=0.23]{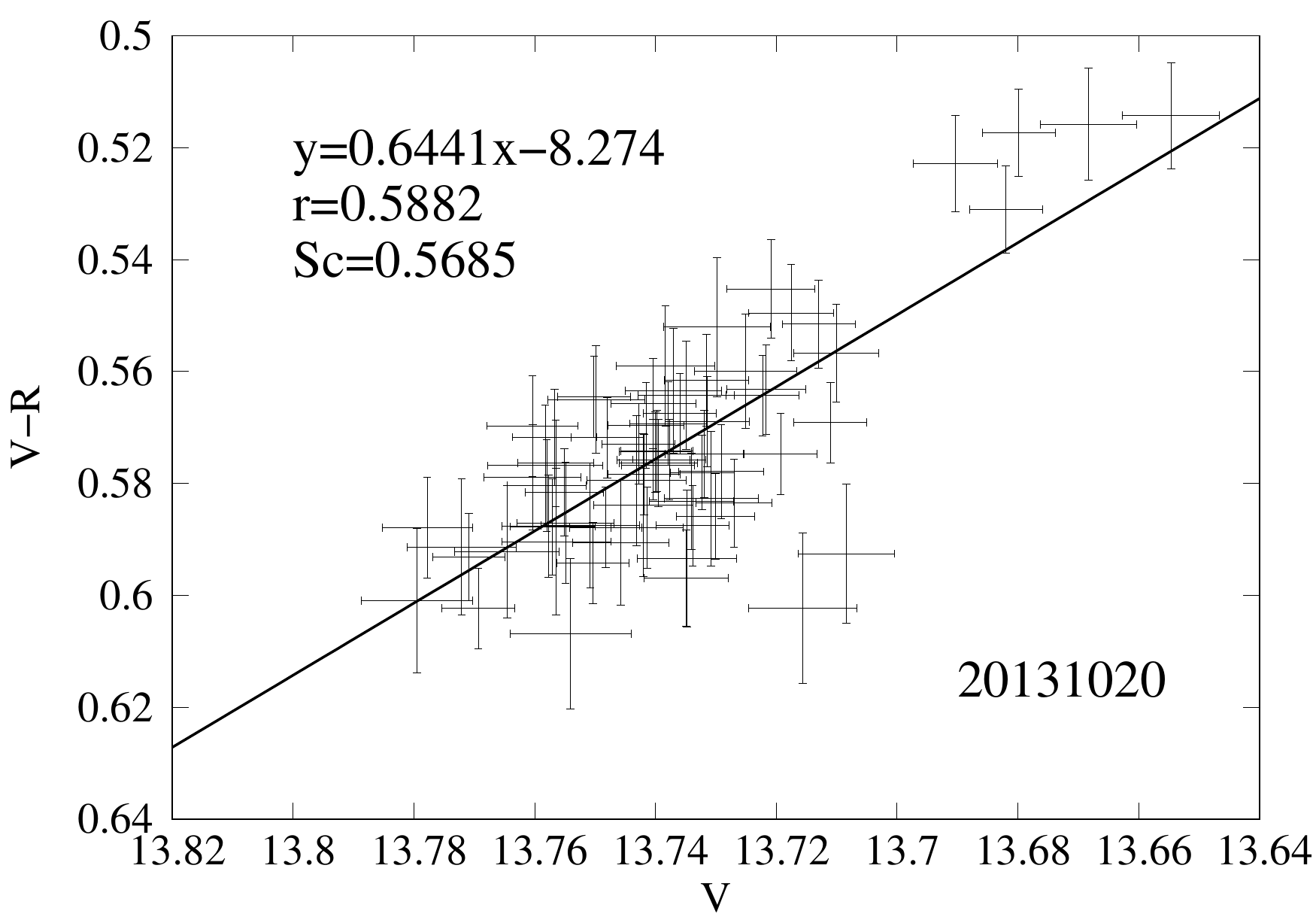}
	\includegraphics[scale=0.23]{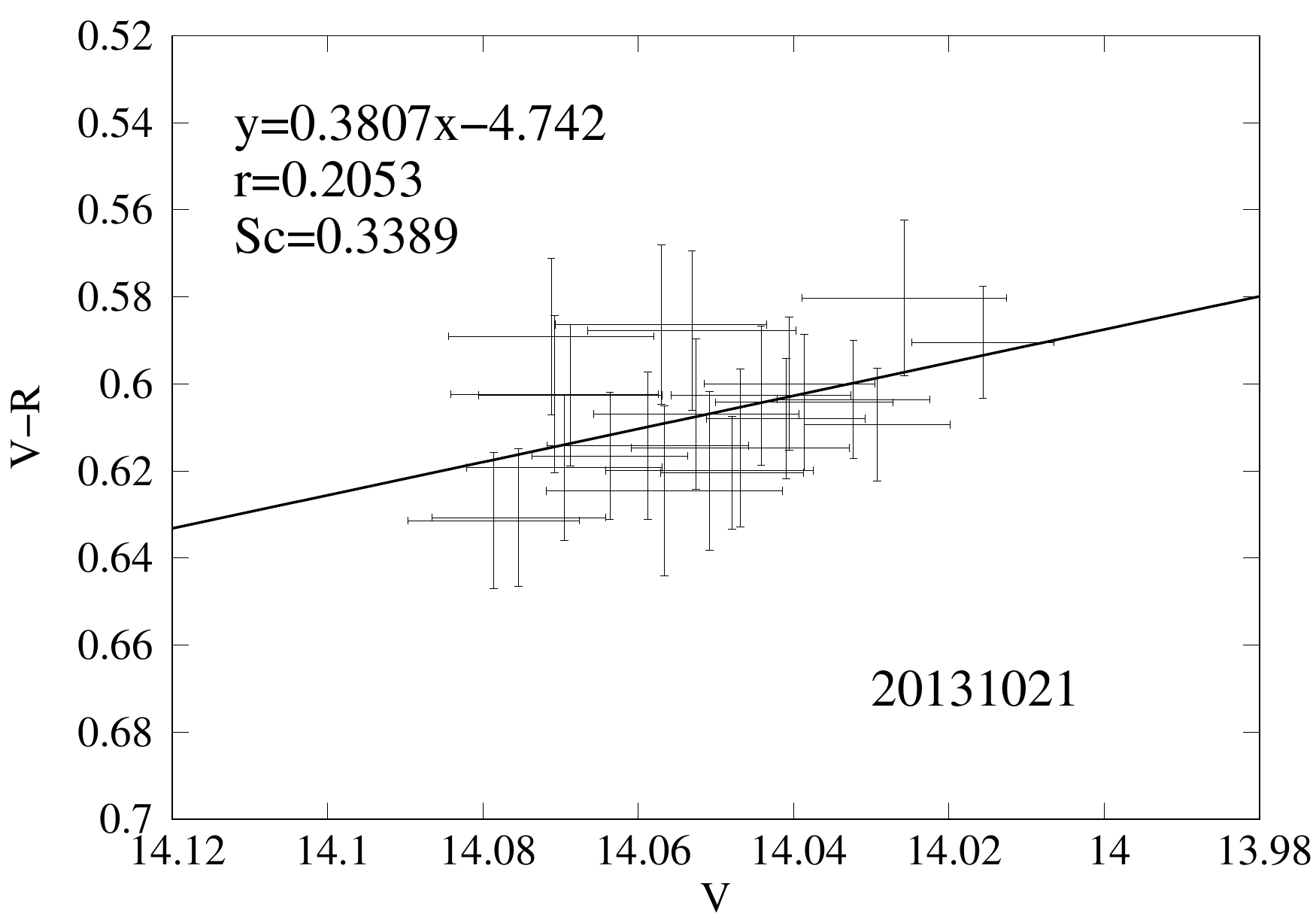}
	\includegraphics[scale=0.23]{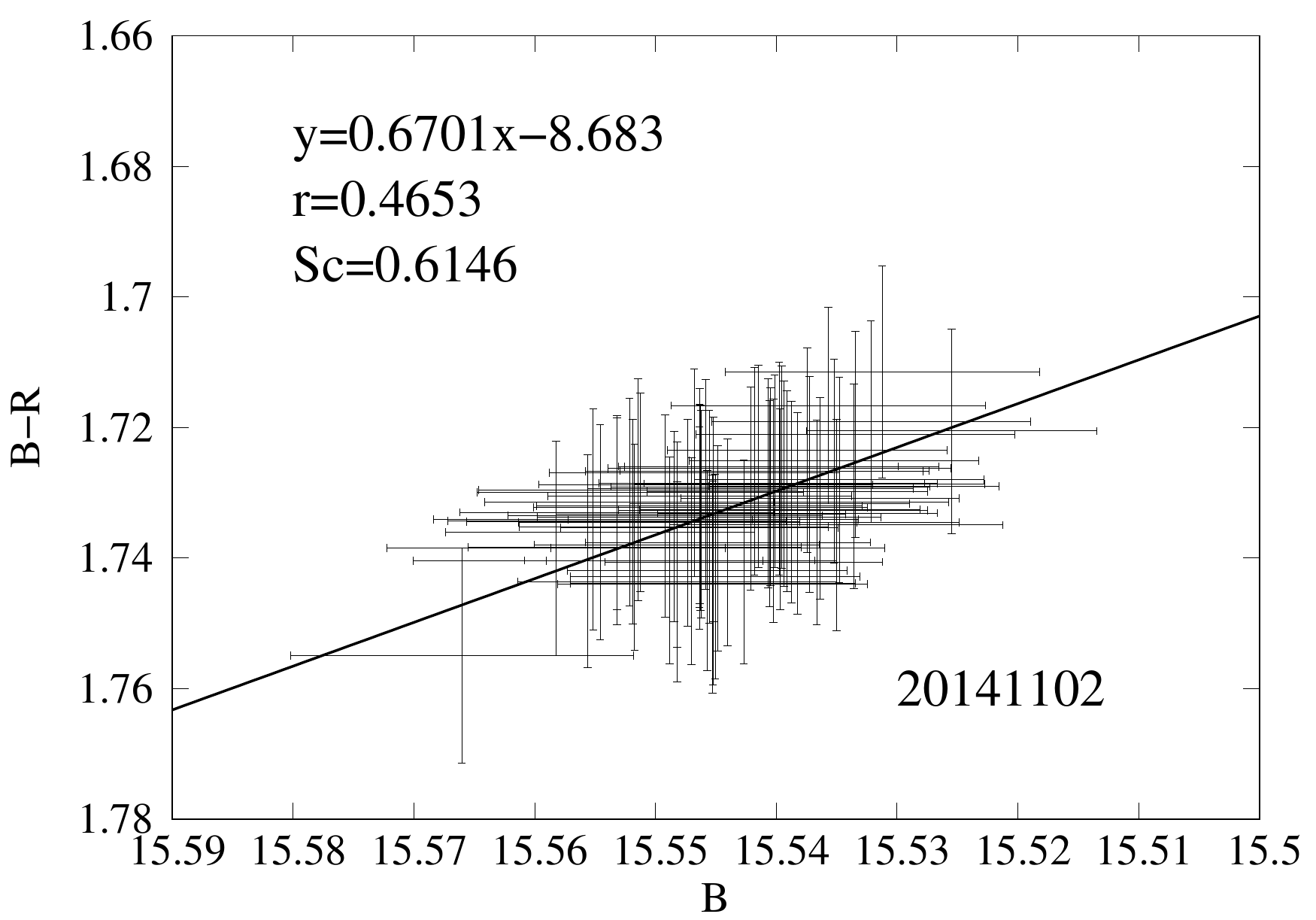}
	\includegraphics[scale=0.23]{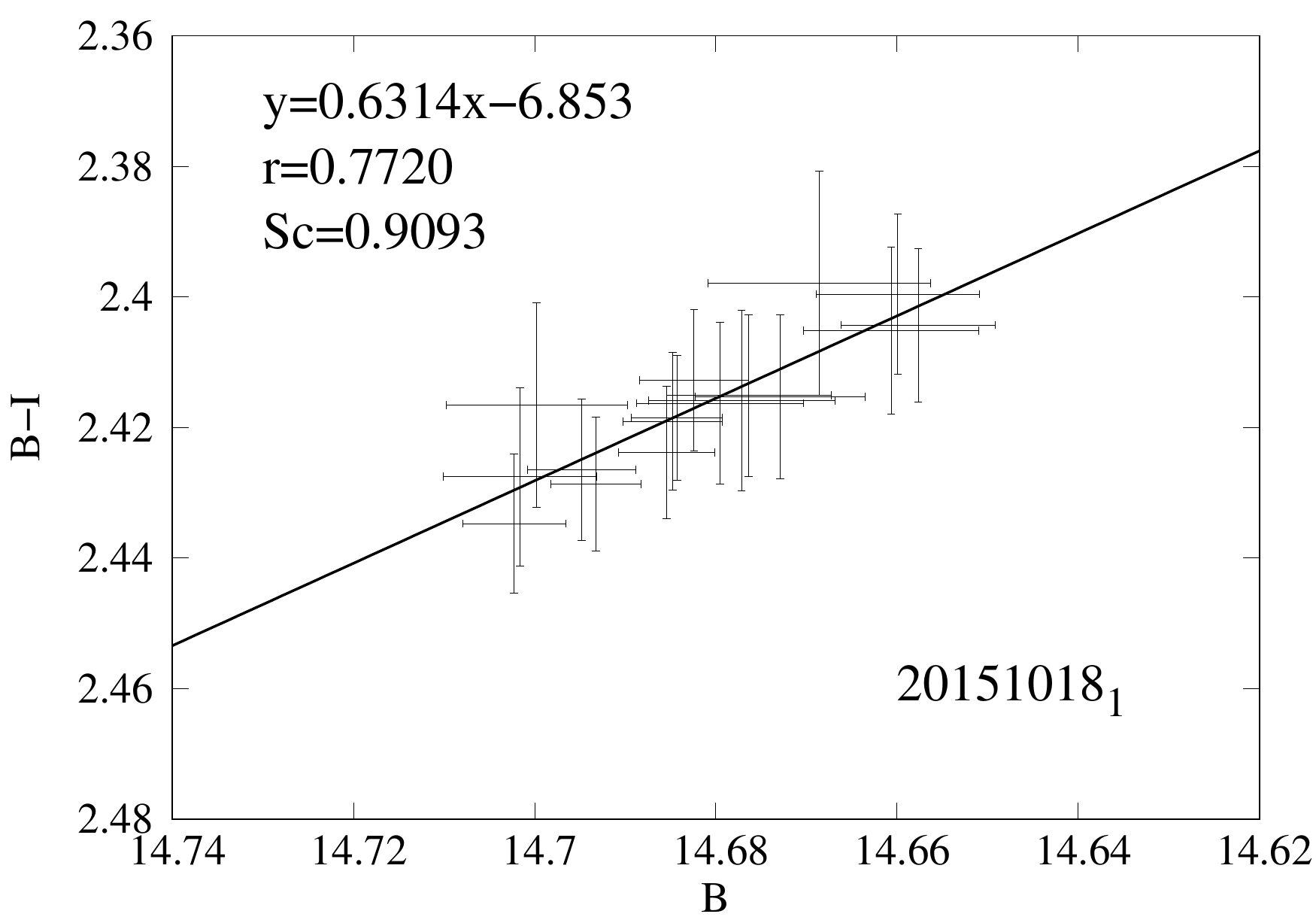}
	\includegraphics[scale=0.23]{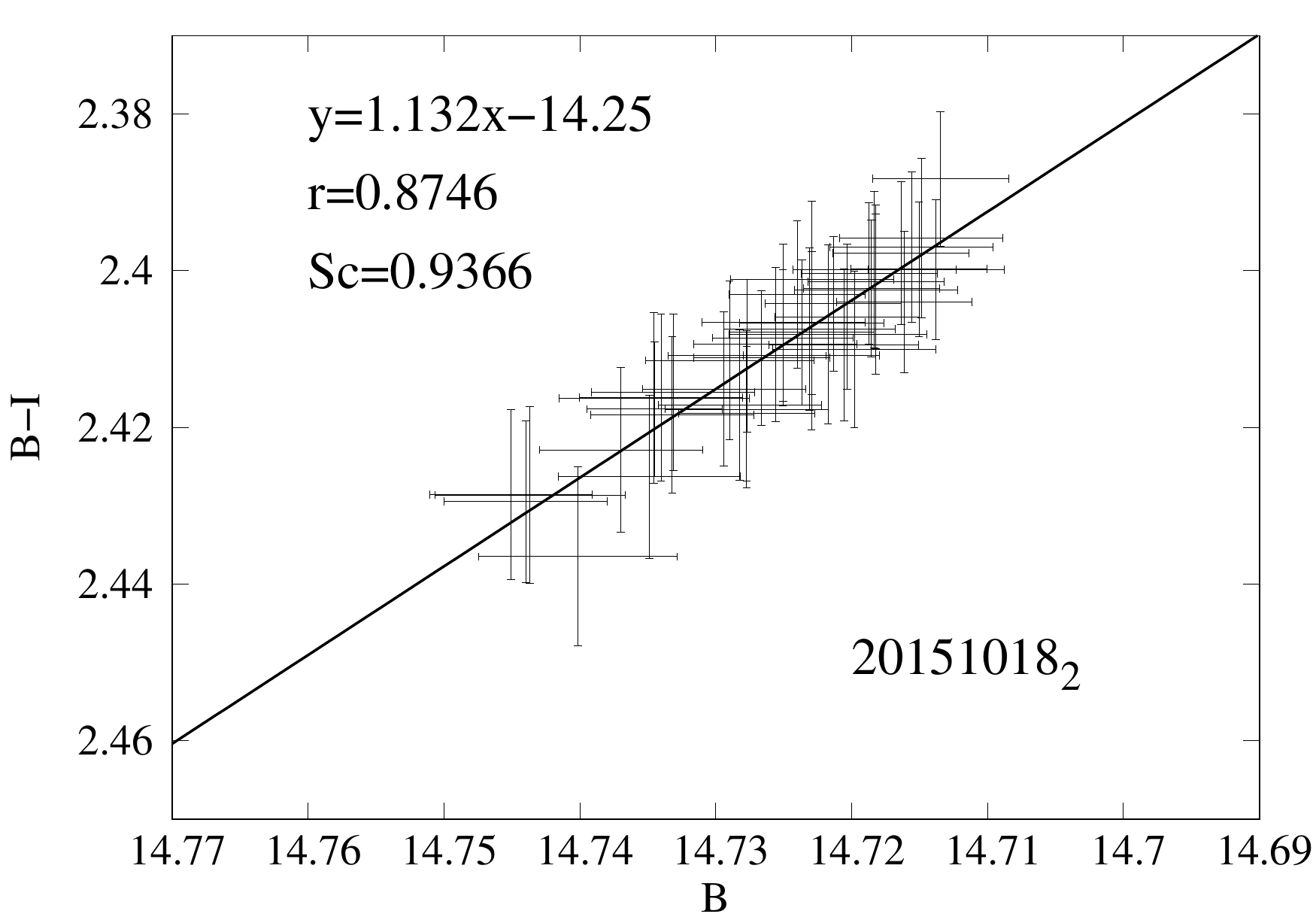}
	\includegraphics[scale=0.23]{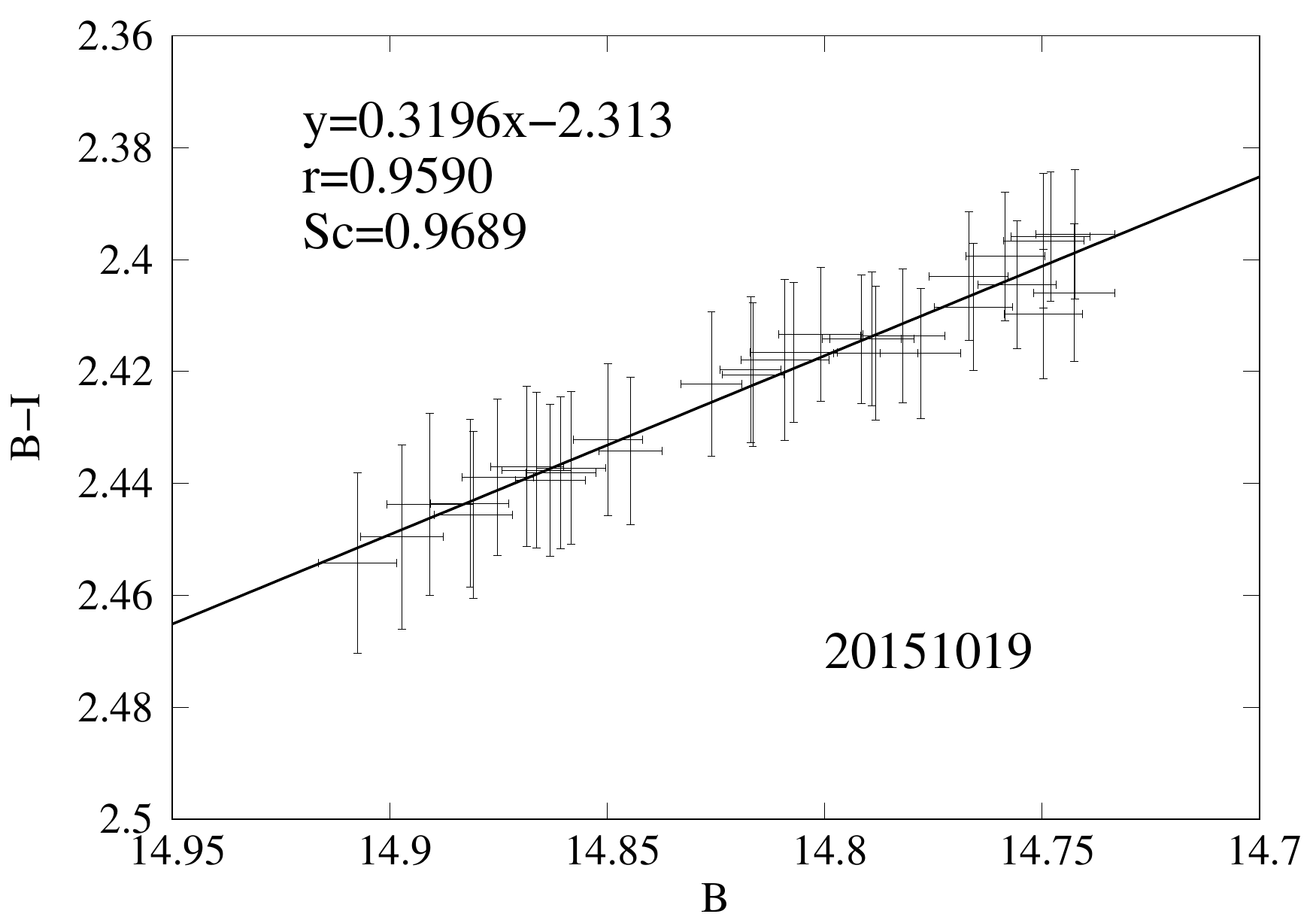}
	\includegraphics[scale=0.23]{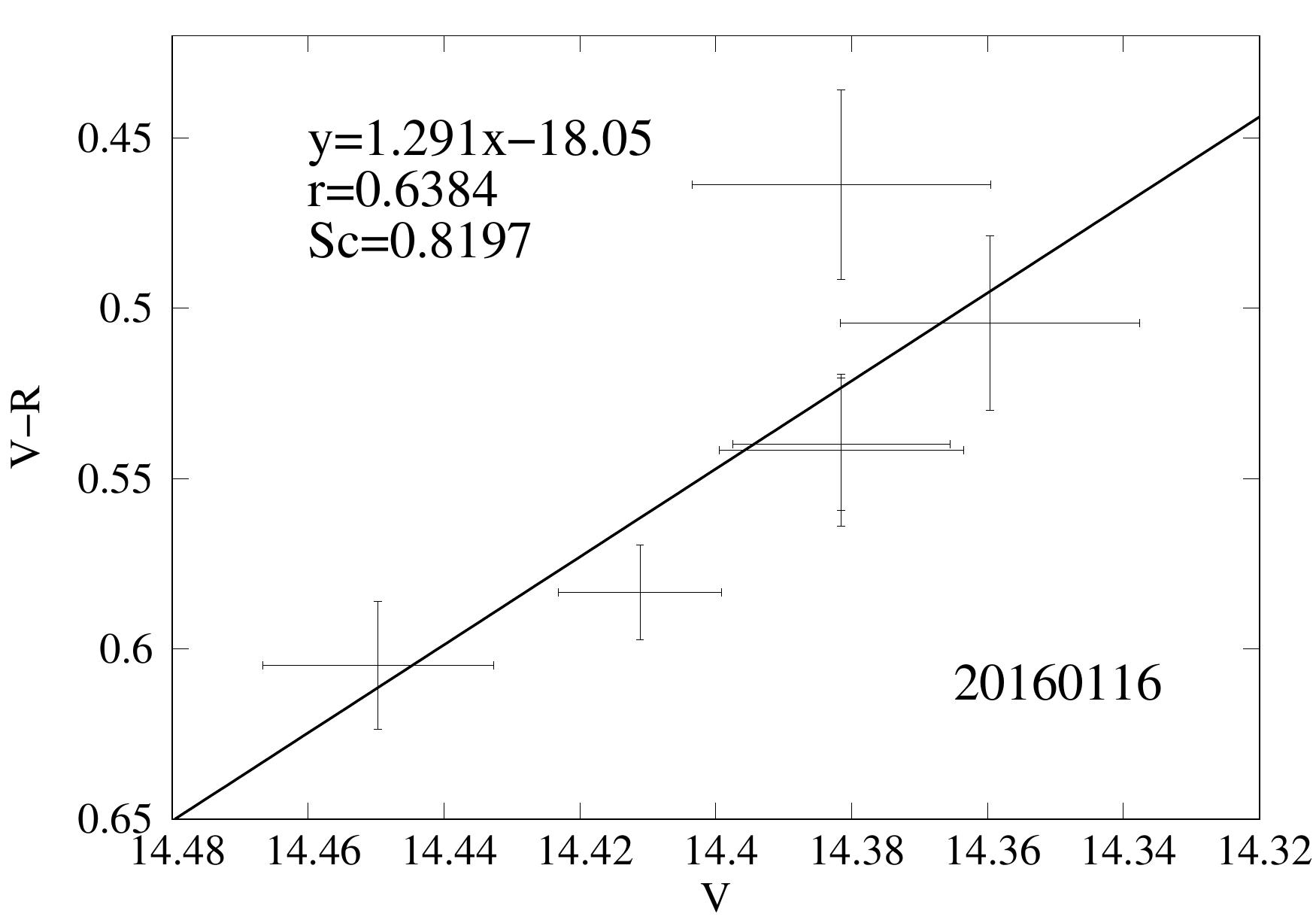}
	\includegraphics[scale=0.23]{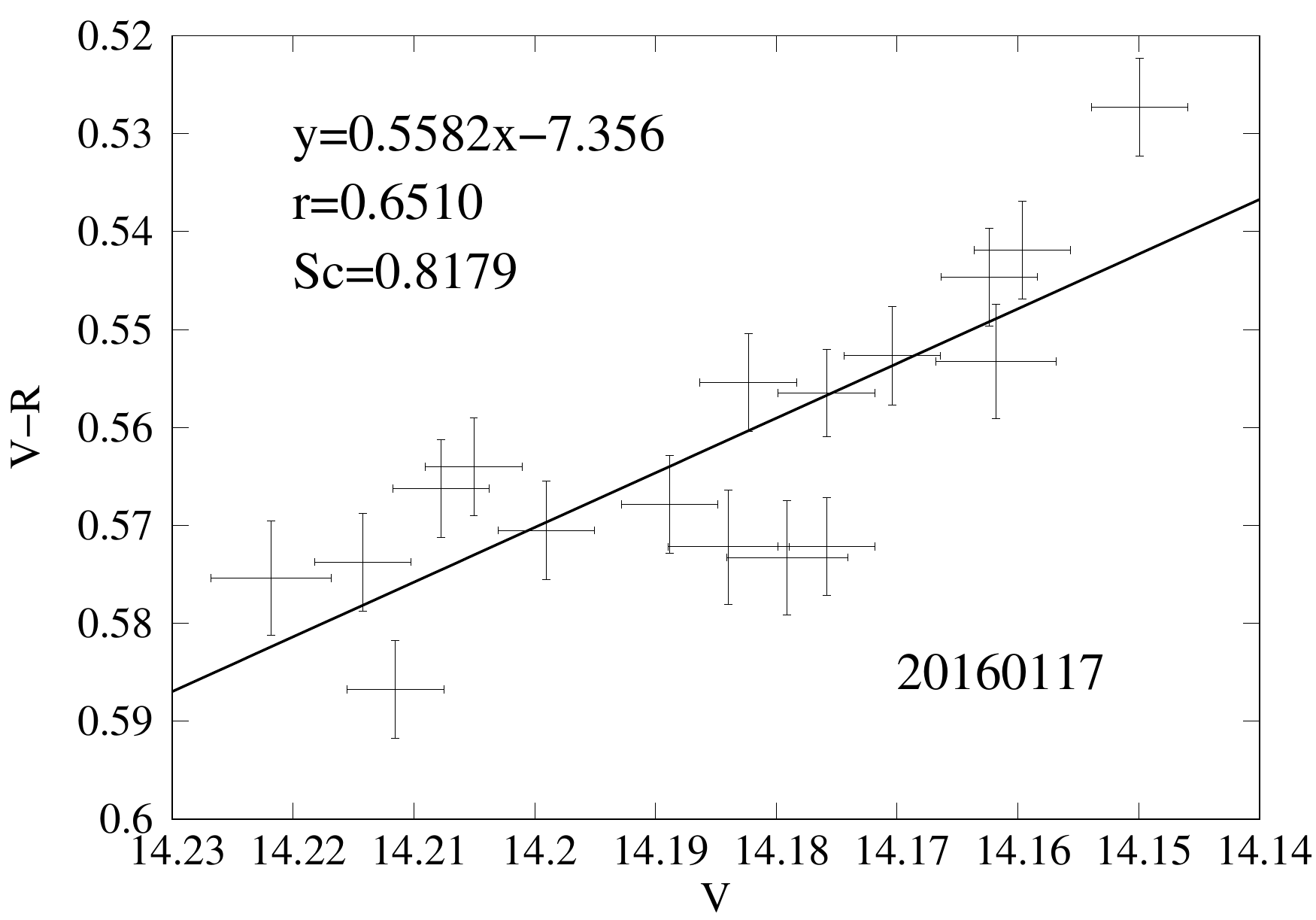}

    \caption{colour-magnitude diagrams of BL Lacertae.}
    \label{fig:CIs}
\end{figure}

\section{Results}

\subsection{Light Curves and Flux Variations}

The light curves we obtained are shown in Fig.~\ref{fig:lightcurves}. We adopted two statistical analysis techniques described in Section 3 to search for variability and the results are listed in Table~\ref{tab:results}. The first column of Table~\ref{tab:results} shows the observation date, while the second column records the telescope. The band and number of the data points are given in the 3rd and 4th columns respectively. The observation durations are listed in the 5th column and the test results of $\chi ^2$ and ANOVA are in the 6th and 7th columns. If $F$ value exceeds the critical value (CV) at the $99\%$ significant level, the null hypothesis that there is no variations will be rejected. BL Lacertae is marked as Y if the variability conditions for each test are satisfied, while N means no variations. The final column is IDV amplitude.

The IDV amplitudes are given by \citet{1996A&A...305...42H}:
\begin{equation}
   A=  \sqrt{(m_{\rm max}-m_{ \rm min})^2-2\sigma^2},
\end{equation}
where $m_{\rm max}$ and $m_{\rm min}$ are the maximum and minimum magnitudes, and $\sigma$ is the standard deviation.

The variations were well correlated in all bands. As can be seen, there are variations on four nights that are detected by both tests. We calculated their amplitudes using the above equation. The maximum amplitude of IDV is $15.85\%$ in the $B$ band on 2015 October 19. The amplitude of IDV is greater in higher energy bands. This has been observed in BL Lacertae \citep{1998AJ....115.2244W,1998A&A...332L...1N,2004NuPhS.132..205N}. The amplitude of variability in different bands is changed during different nights. The comparison results of two tests indicate that ANOVA shows efficient detection of IDV with small flares. Several small flares can be seen in the light curves on four of these nights, which can be claimed as the IDV by ANOVA test. We interpreted the observed flares in terms of the model consisting of individual synchrotron pulses in another section below.

\begin{figure}
	\centering
	\includegraphics[scale=0.23]{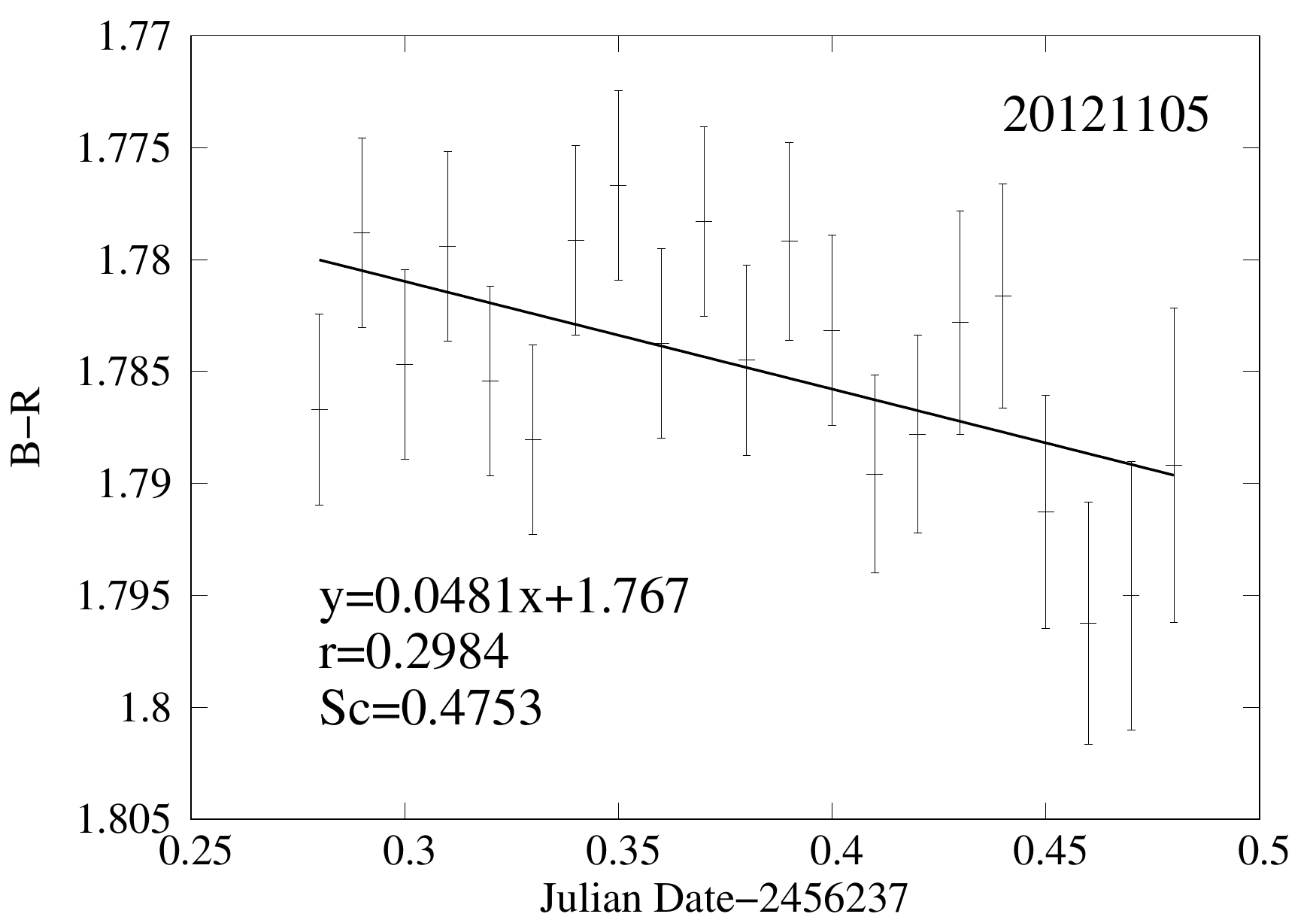}
	\includegraphics[scale=0.23]{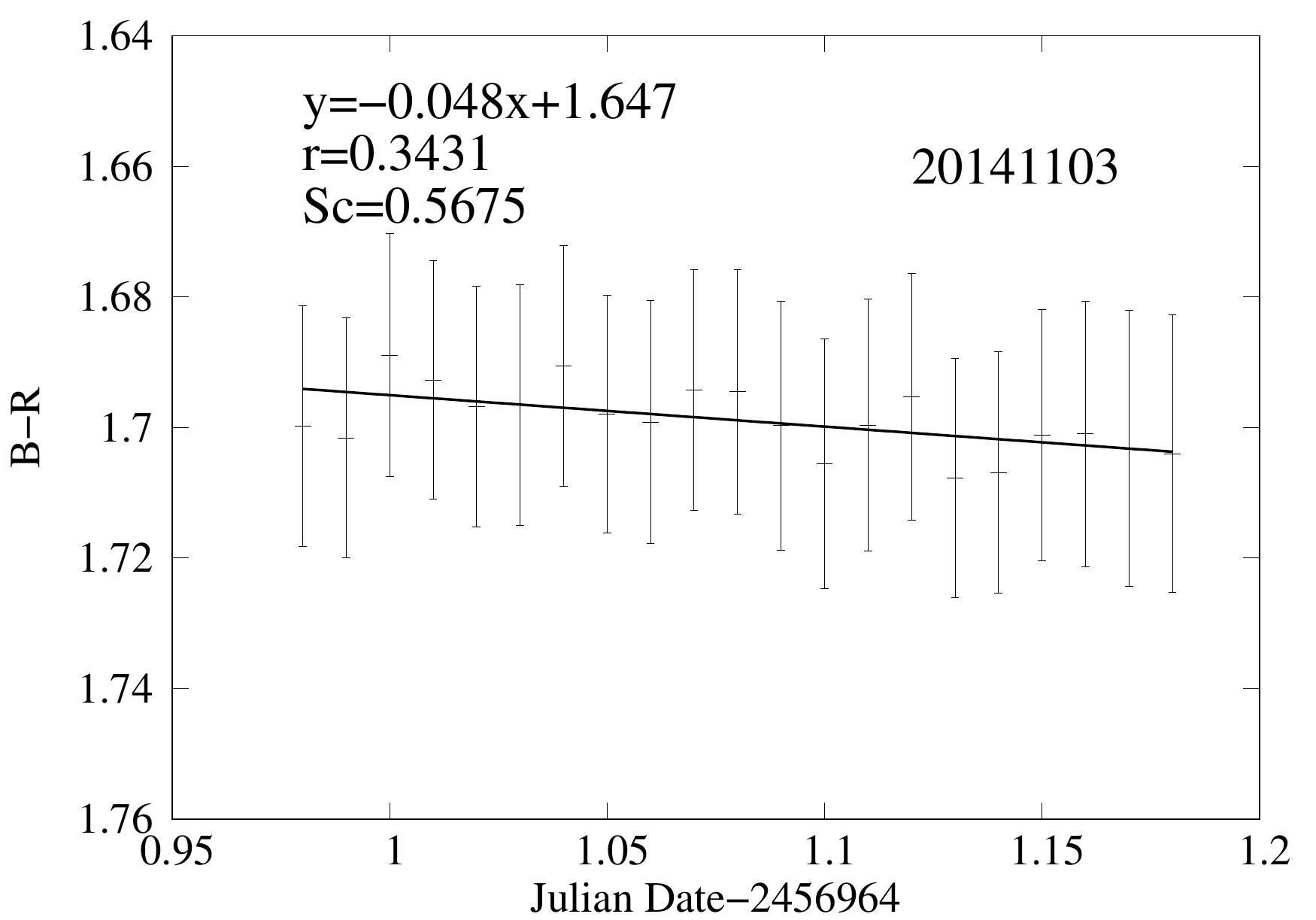}
	\includegraphics[scale=0.23]{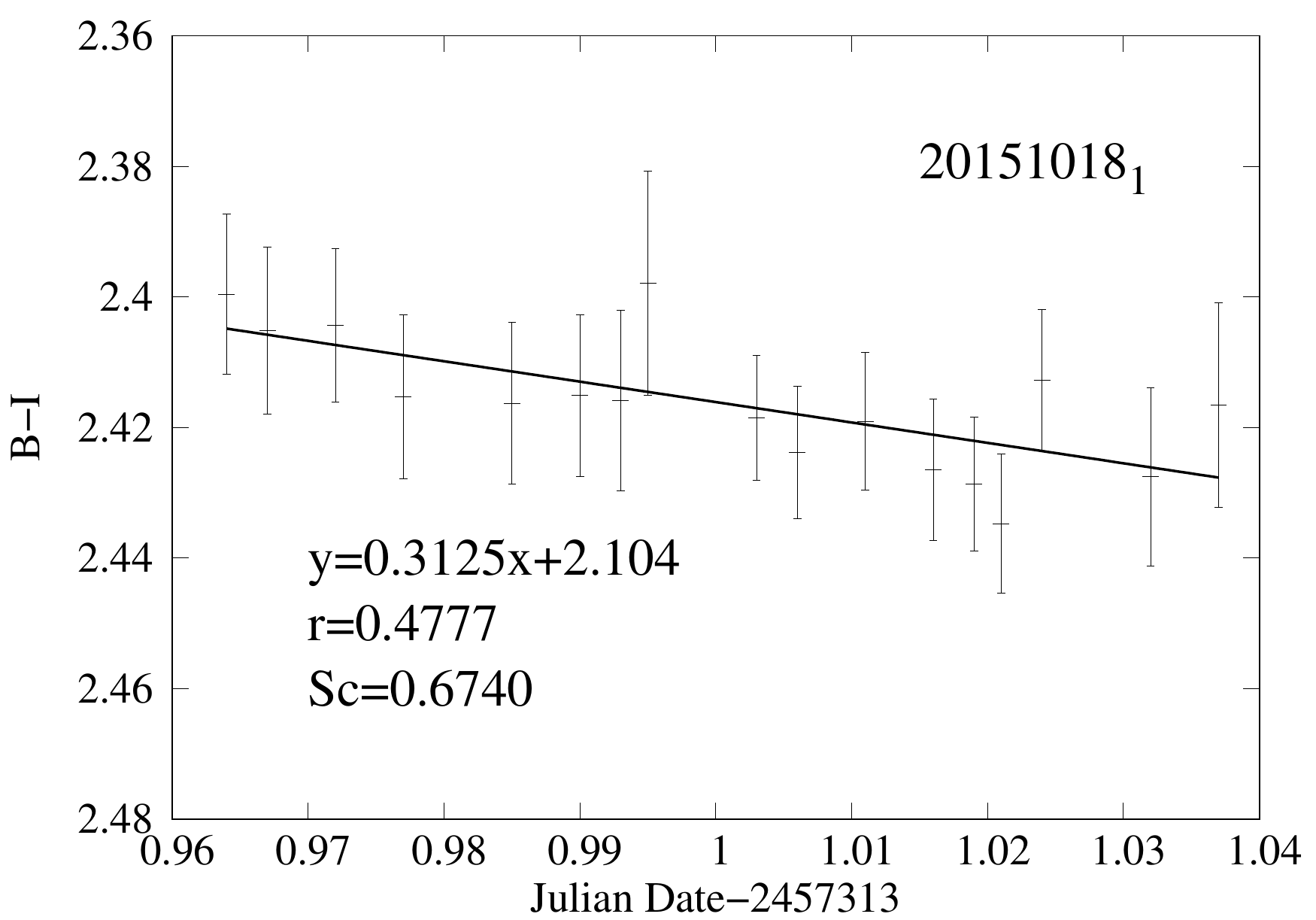}
	\includegraphics[scale=0.23]{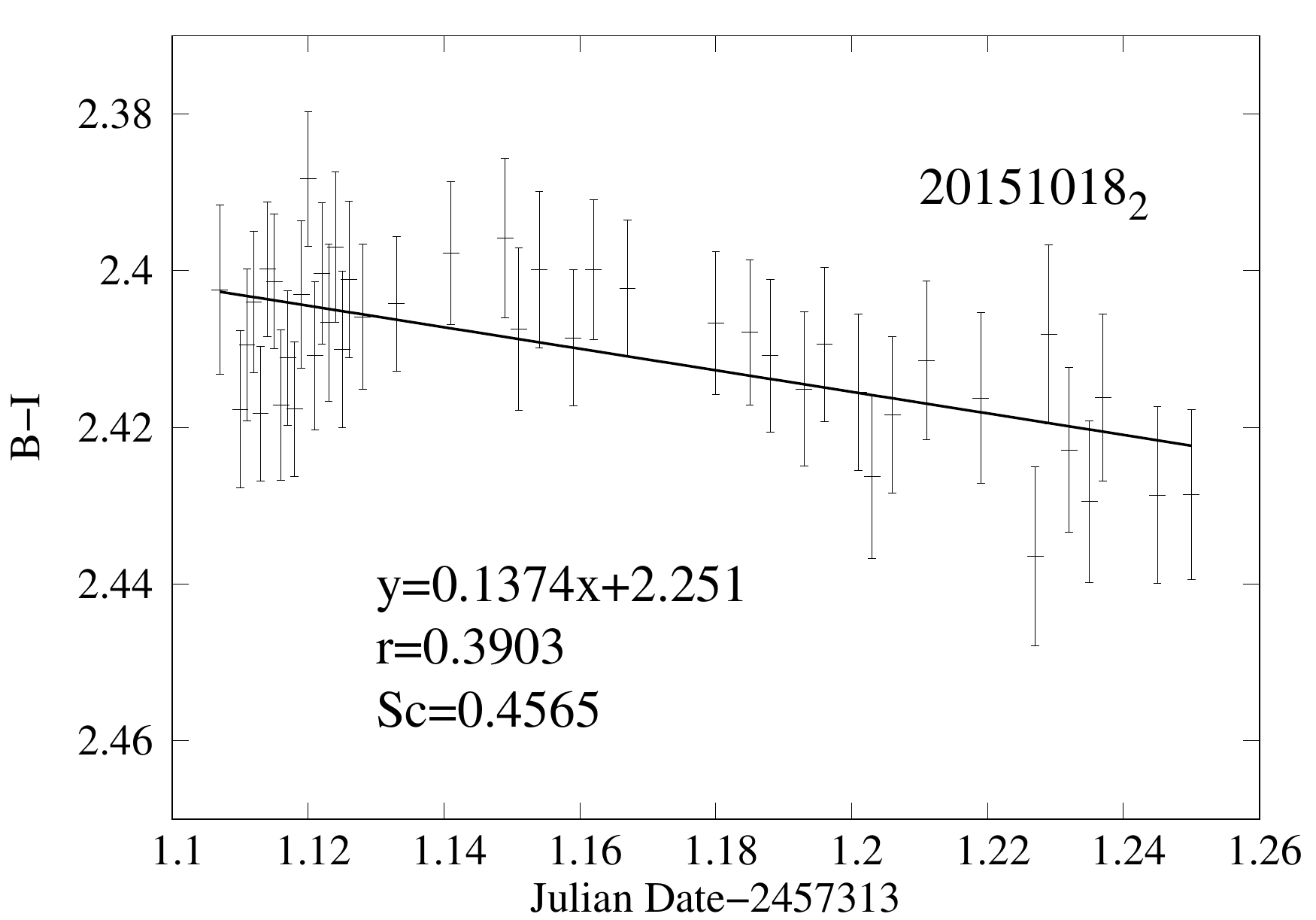}
	\includegraphics[scale=0.23]{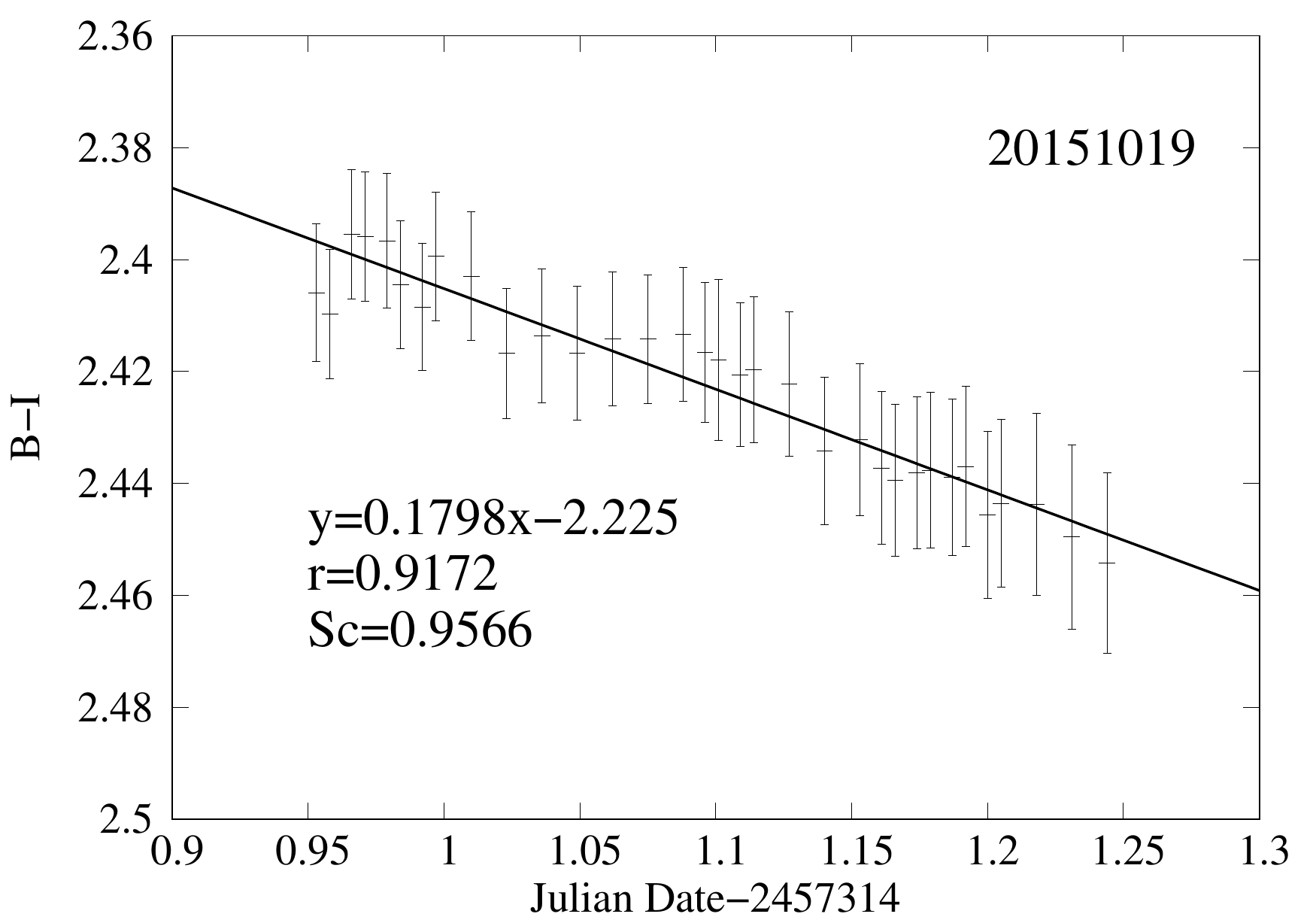}
	\includegraphics[scale=0.23]{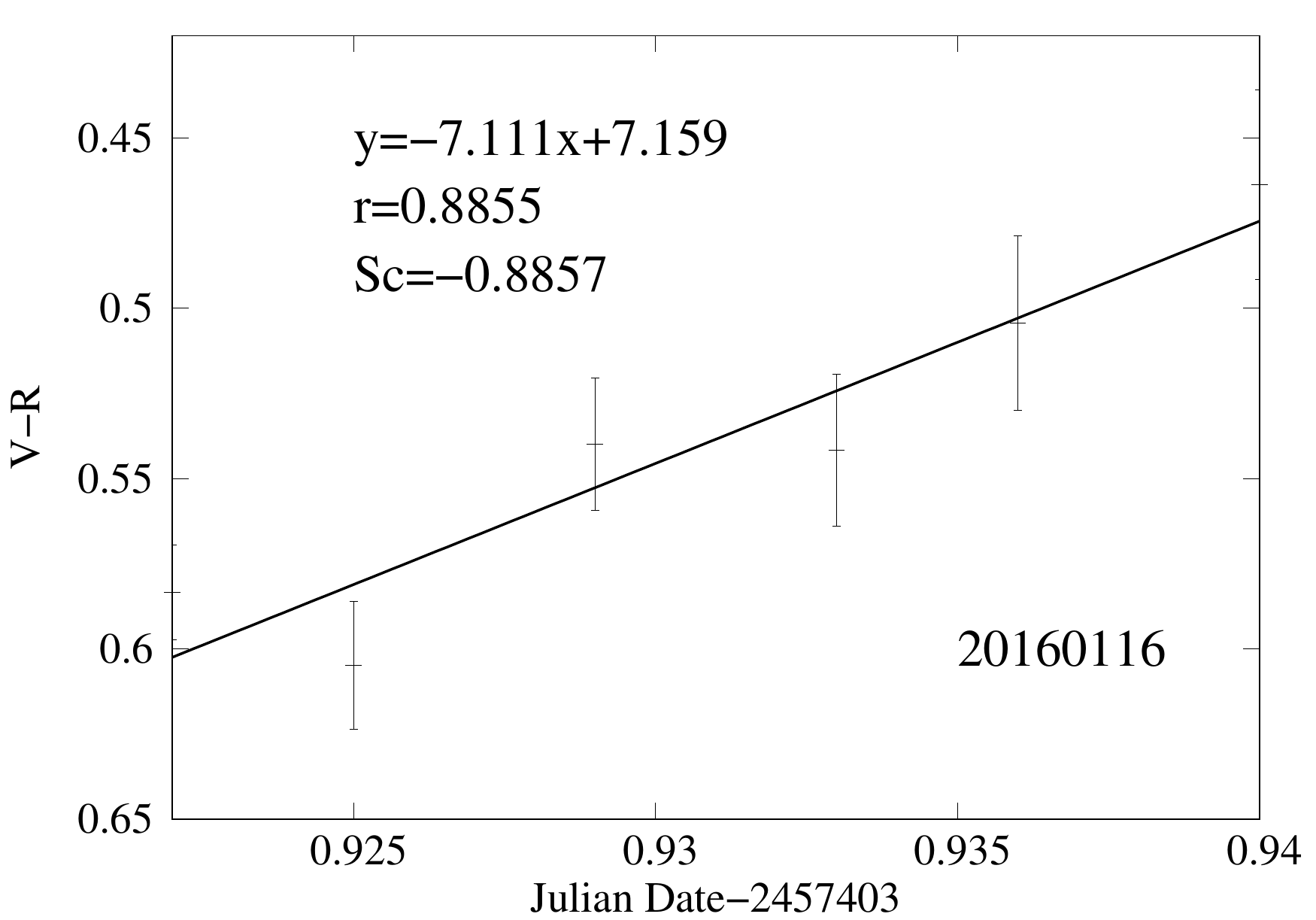}
	\includegraphics[scale=0.23]{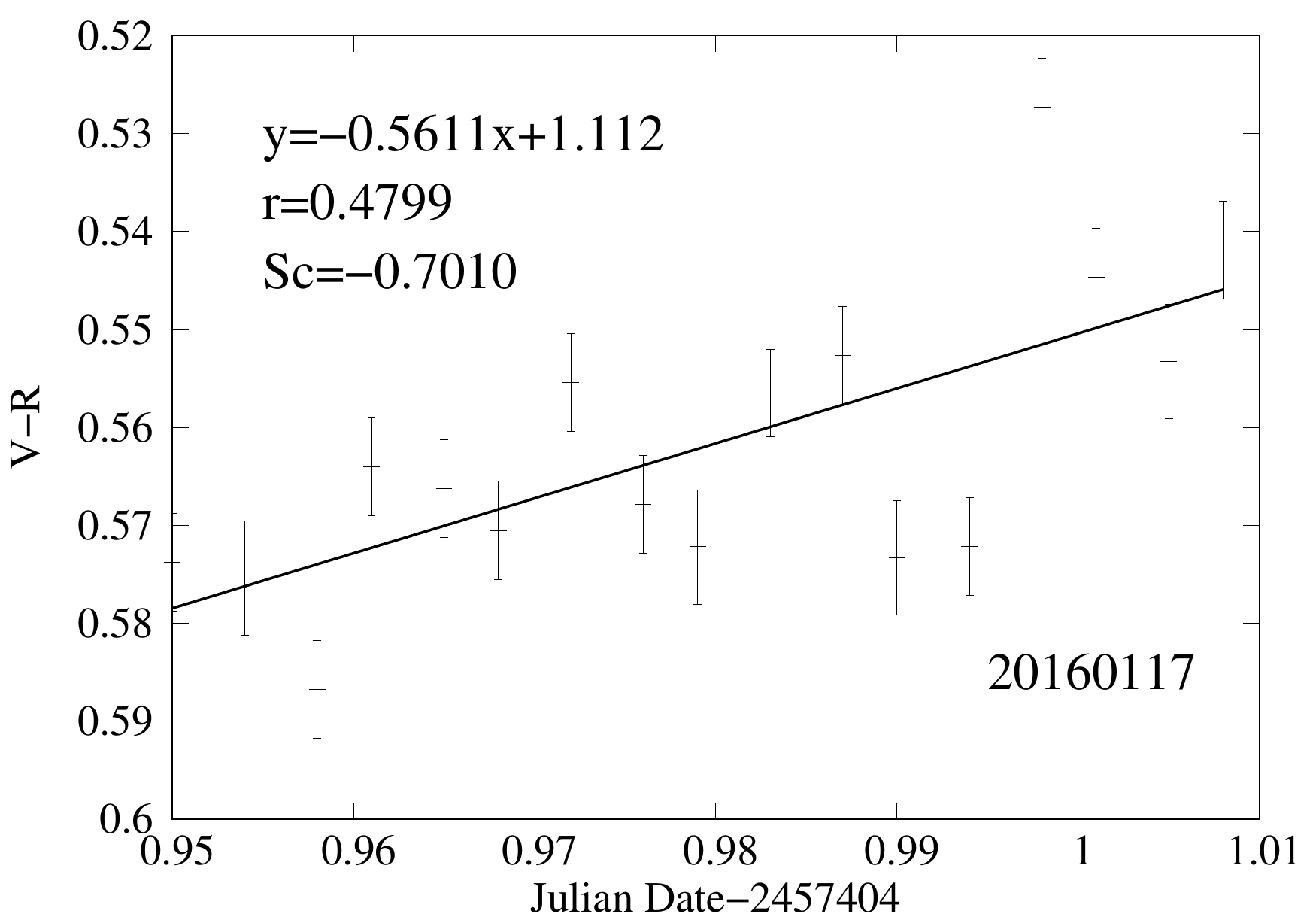}

      \caption{colour index versus intraday time-scale.}
      \label{fig:colourtime}
\end{figure}

\begin{figure}

	\includegraphics[scale=0.45]{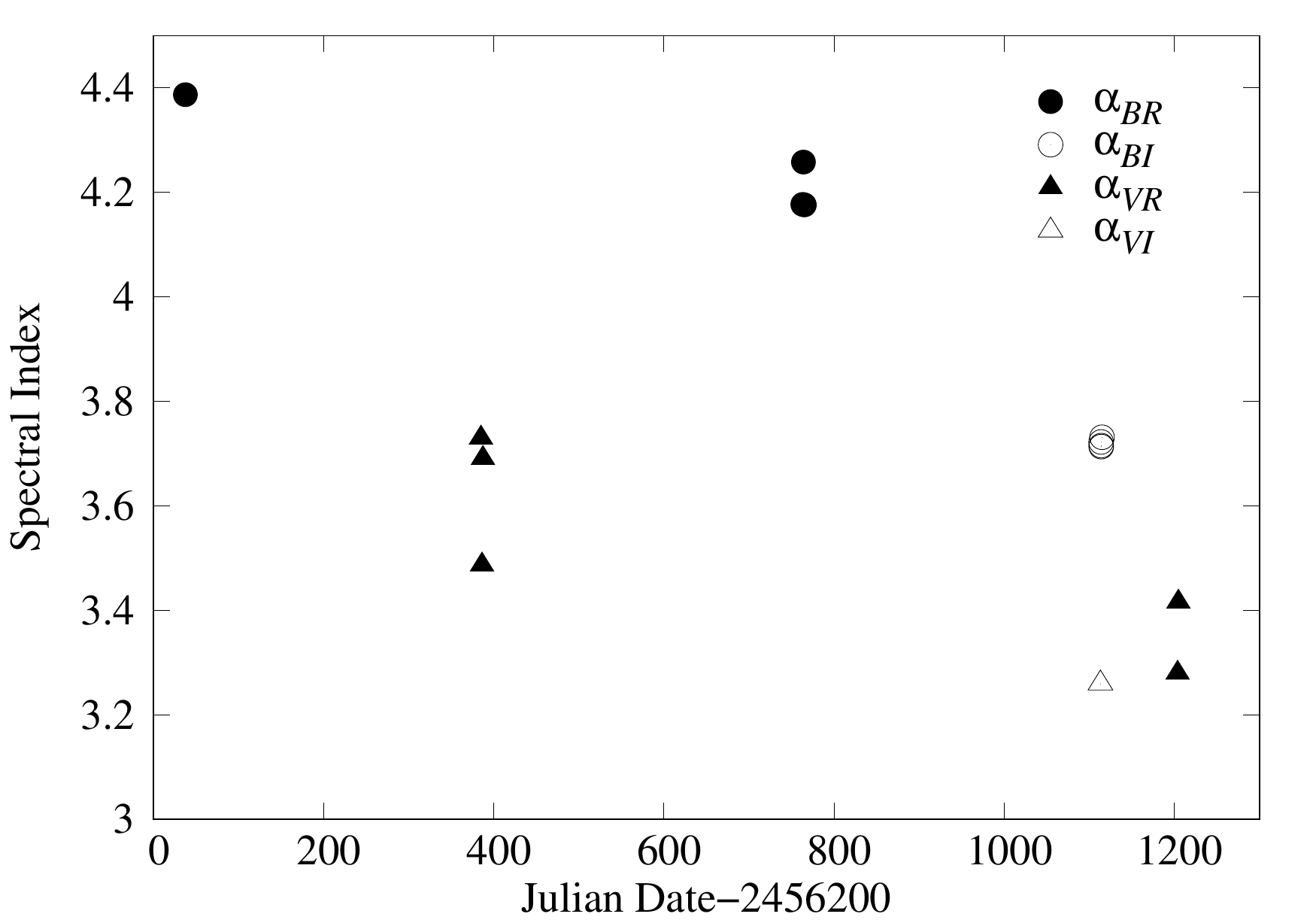}
    \caption{Variation of average optical spectral index versus time covering the entire observation period for the target.}
    \label{fig:sitime}
\end{figure}

\subsection{colour behaviour}

We investigated the colour behaviour with respect to the brightness of BL Lacertae for each separate night. The colour indices of $B-R, V-R, V-I$ and $B-I$ are calculated by using the almost simultaneous $B, V, R$ and $I$ magnitudes. Because of the existence of gaps in the light curves on 2015 October 18, we plotted two separate diagrams and the results are displayed in Fig.~\ref{fig:CIs}. We fitted the colour-magnitude diagrams with a linear model (where r is the correlation coefficient) and calculated the Spearman correlation coefficient (Sc). Three of them were left out since they were not able to meet the conditions of the $F$-test after linear regression analysis.

The source exhibits a bluer-when-brighter (BWB) trend that has been reported previously \citep{1970ApJ...159L..99R,1998A&A...339..382S,2003ApJ...590..123V,2006MNRAS.366.1337S,2007A&A...470..857P,2011PASJ...63..639I,2015MNRAS.452.4263G,2015A&A...573A..69W} and unambiguously confirmed as a universal aspect in blazars by \citet{2011PASJ...63..639I}. Especially on 2015 October 19, the Sc value reaches up to 0.9689, indicating the significant linear correlation between colour index and magnitude. Too few data points on several observations, particularly on 2016 January 16, made the fitting unreliable. The Sc values are less than 0.6 on three nights that all have flares. The existence of flares may have the opposite effect on the colour behaviour of BL Lacertae. \citet{2015MNRAS.452.4263G} interpreted the weakening of the colour-magnitude correlations as the superposition of many distinct new variable components. \citet{galaxies4030015} proposed a new definition of microvariability as the short term oscillations distinct from the linear variability.
\citet{2015A&A...573A..69W} argued that the observed BWB behaviour was intrinsic to the jet emission regions. 

The colour trend in blazars on intraday time-scales can help us investigate the origin of blazar emissions. Fig.~\ref{fig:colourtime} shows the plots of each colour index against intraday time-scale with fitting lines. The best-fitting values of slope, intercept, r, Sc, and null hypothesis probability are listed in Table~\ref{tab:colourfits}. A positive slope implies significant positive correlation between colour index and time when both r and Sc are greater than 0.9. We also gave the F-test results and the critical values after linear regression analysis. Since some of them were not able to meet the conditions of the $F$-test, we did not plot them in Fig.~\ref{fig:colourtime}. The average colours $\rm <CI>$ and the corresponding spectral indices $\rm <SI>$ are given in Table~\ref{tab:colourfits} as well. According to \citet{2015A&A...573A..69W}, the average spectral indices are derived simply as

\begin{equation}
   <\alpha_{\rm A B}>=\frac{0.4<A-B>}{log(\nu_{\rm A}/\nu_{\rm B)}} ,
\end{equation}
where $A$ and $B$ stand for different bands, and $\nu_{\rm A}$ and $\nu_{\rm B}$ are effective frequencies of the respective bands \citep{1998A&A...333..231B}.
The spectral indices varied slightly as shown in Fig.~\ref{fig:sitime}. 
The spectral indices of $\alpha_{BR}$, $\alpha_{BI}$ and $\alpha_{VR}$ changed by only 0.21, 0.02 and 0.45, respectively. The $\alpha_{VI}$ even remained unchanged due at least partly to the few data points.
The accretion disc radiation is expected to be overwhelmed by that from the strongly Doppler-boosted jets, so the observed spectral variations in  blazars cannot be explained by the accretion disc. The relatively steep spectral indices indicate strong synchrotron emission from the blazar jet and small accretion disc contribution \citep{2016MNRAS.455..680A}.

\subsection{Cross-correlation analysis and time lags}

We performed the correlation analysis to search for the possible inter-band time lags by using two cross-correlation methods. The first one is the z-transformed discrete correlation functions (ZDCFs) method \citep{1997ASSL..218..163A}. ZDCF deals with under-sampled light curves and divides all observation points into equal bins. It uses Fisher's z-transform to stabilize the highly  skewed distribution of the correlation coefficient. The Gaussian fitting (GF) is made to the central ZDCF results. Meanwhile, we try another way to measure the lags and errors by interpolated cross-correlation function (ICCF) method \citep{1987ApJS...65....1G}. The error was estimated with a model-independent Monte Carlo method, and the lag was taken as the centroid of the cross-correlation functions that were obtained with a large number of independent Monte Carlo realizations. This is the flux-randomization/random-subset selection (FR/RSS) approach described by \citet{1998PASP..110..660P,2004ApJ...613..682P}. Five thousand independent Monte Carlo realizations were performed on each light curve.

\begin{figure}
	\includegraphics[scale=0.45]{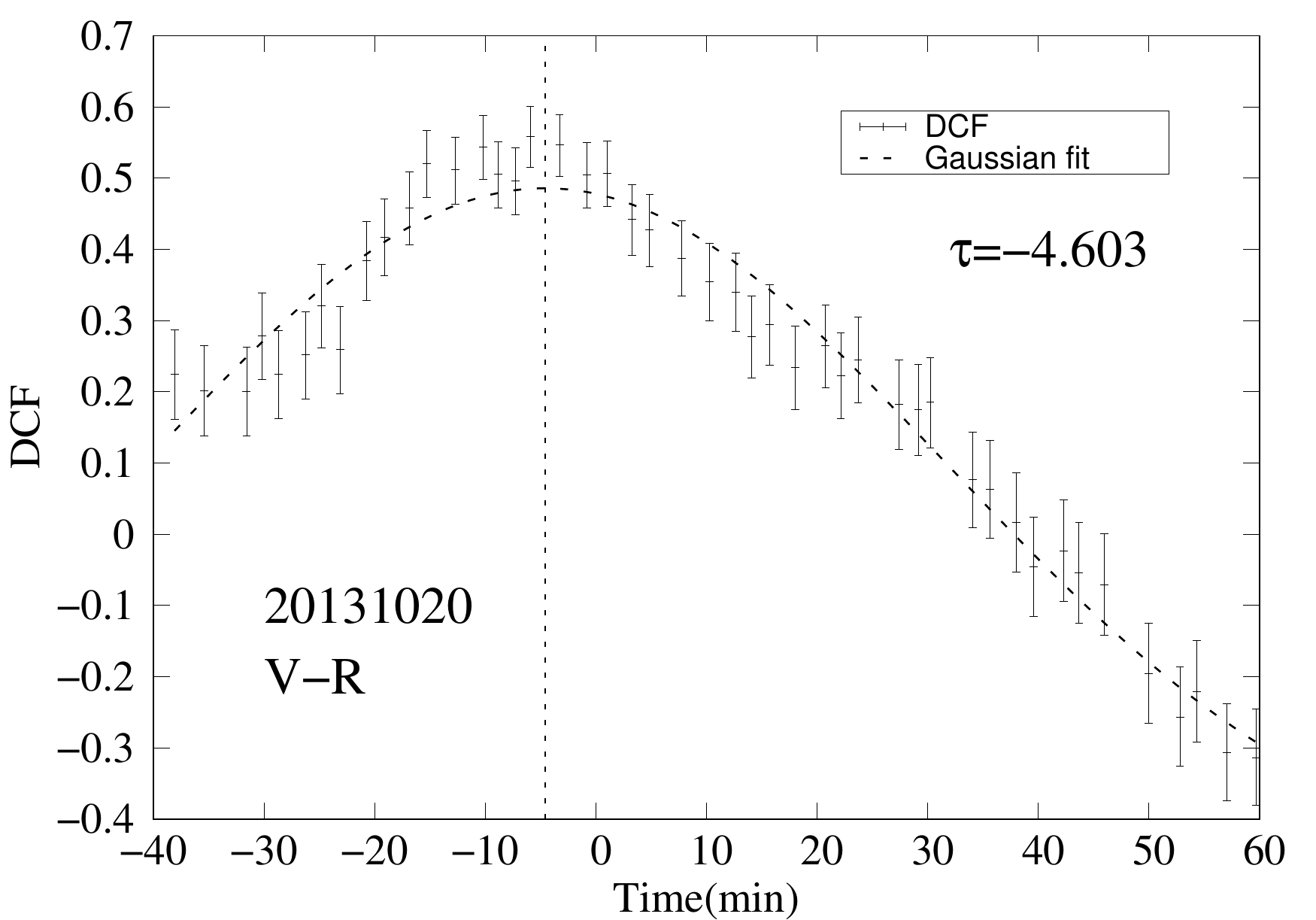}	
	
    \caption{ZDCF correlation and fitting result on 2013 October 20. The dashed line shows Gaussian fitting to the points, and the peak is marked with the vertical dotted line. $\tau$ gives the lag result.}
    \label{fig:lag}
\end{figure}

One problem in the GF is that it usually underestimates the error for time delay \citep{2012AJ....143..108W}. The results for ZDCF+GF are only for reference. The FR/RSS lags have significance lower than $3\sigma$ except for the $V-R$ lag on 2013 October 20. On that night, the variability in the $R$ band led that in the $V$ band by 11.8 min. The correlation analysis plot is displayed in Fig.~\ref{fig:lag}. Date and the correlated passbands are given at left side. The peak of the Gaussian profile (the dashed line) is marked with a vertical dotted line. A negative lag ($\tau$) means that the later variation leads the former one. Because of large errors, no time delays were found on other nights.
Since the data were binned at 3-min intervals and we used the binned data to estimate the time lags, so the result of 11.8 mins should be reasonable.

\citet{2012AJ....143..108W} discussed the possible key factors that determine the detectability of the optical time lags. Because the $V$ and $R$ bands are too close, the starting time of the variations would be almost the same. It is difficult to detect the time lags between optical bands. The similar flare structures are conducive to the study of time lag detection. 
Maximizing the number of data points and using integration times will lead to good signal-to-noise \citep{1988AJ.....95..247H}. 
The exposure times for this source were too short to get good signal-to-noise data in 2014. 
A same temporal resolution of about 1 min in different optical bands is favourable in the time-lag detection of the BL Lacertae.

\begin{figure}
	\centering
	\includegraphics[scale=0.36]{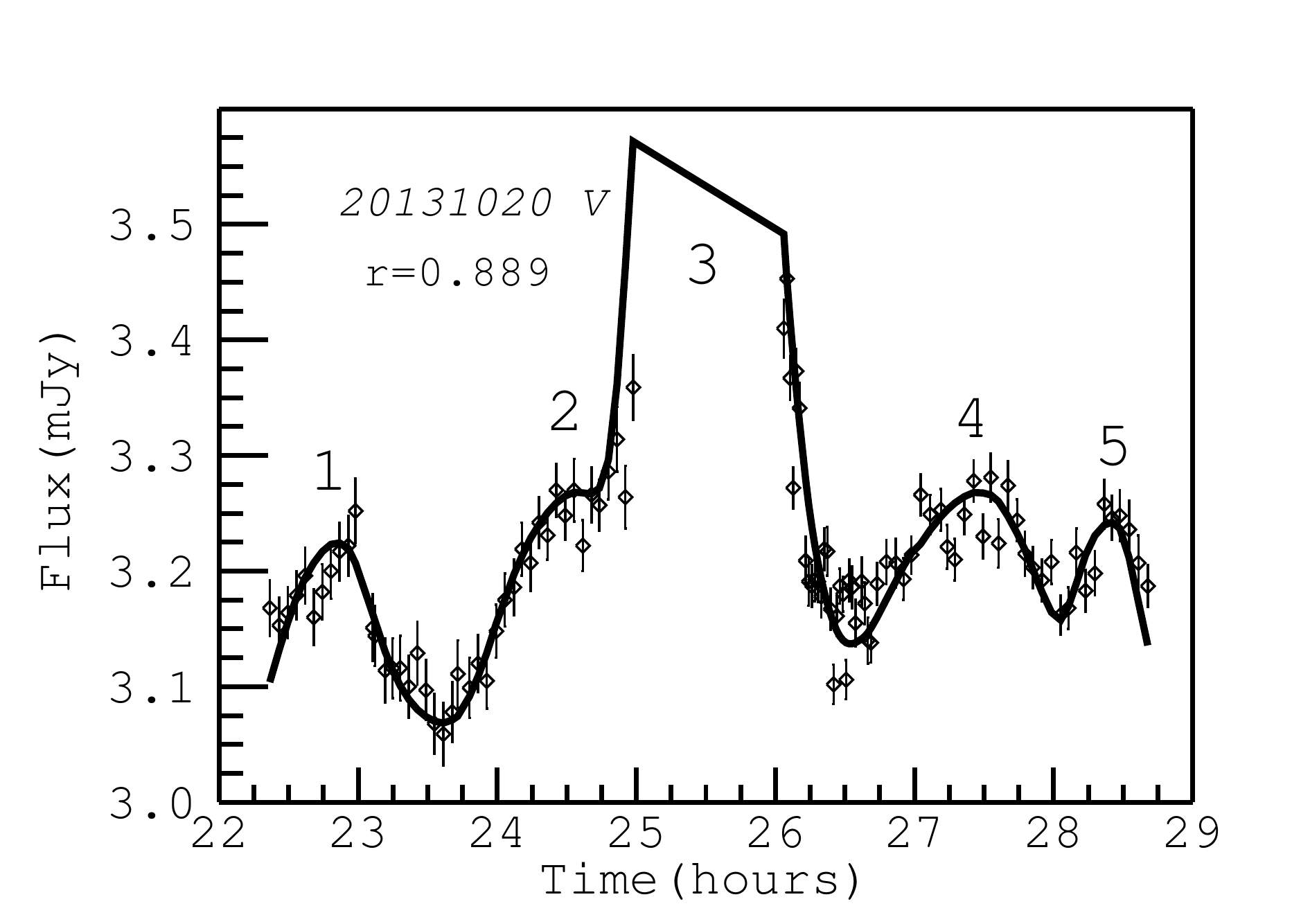}
	\includegraphics[scale=0.36]{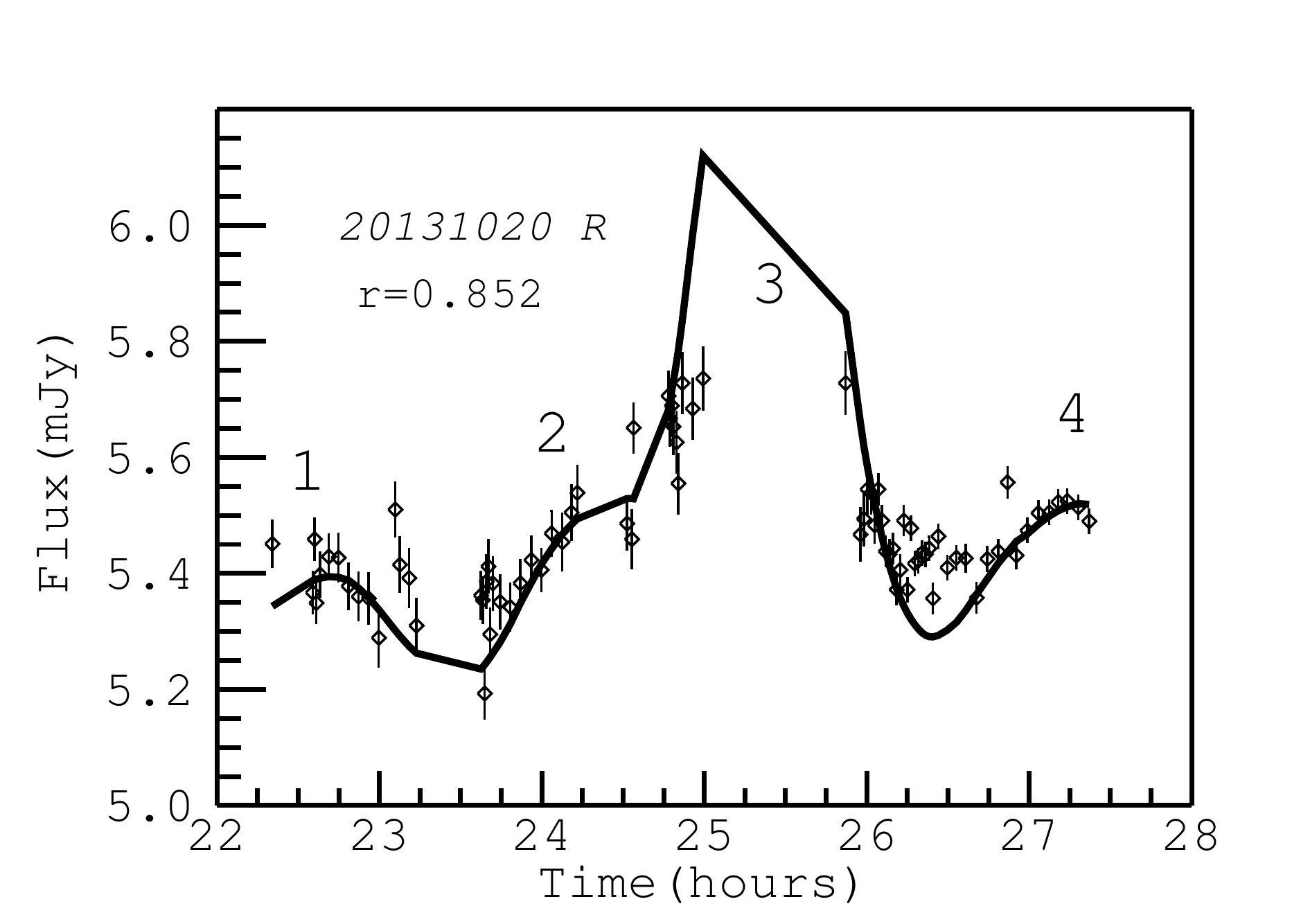}
	\includegraphics[scale=0.36]{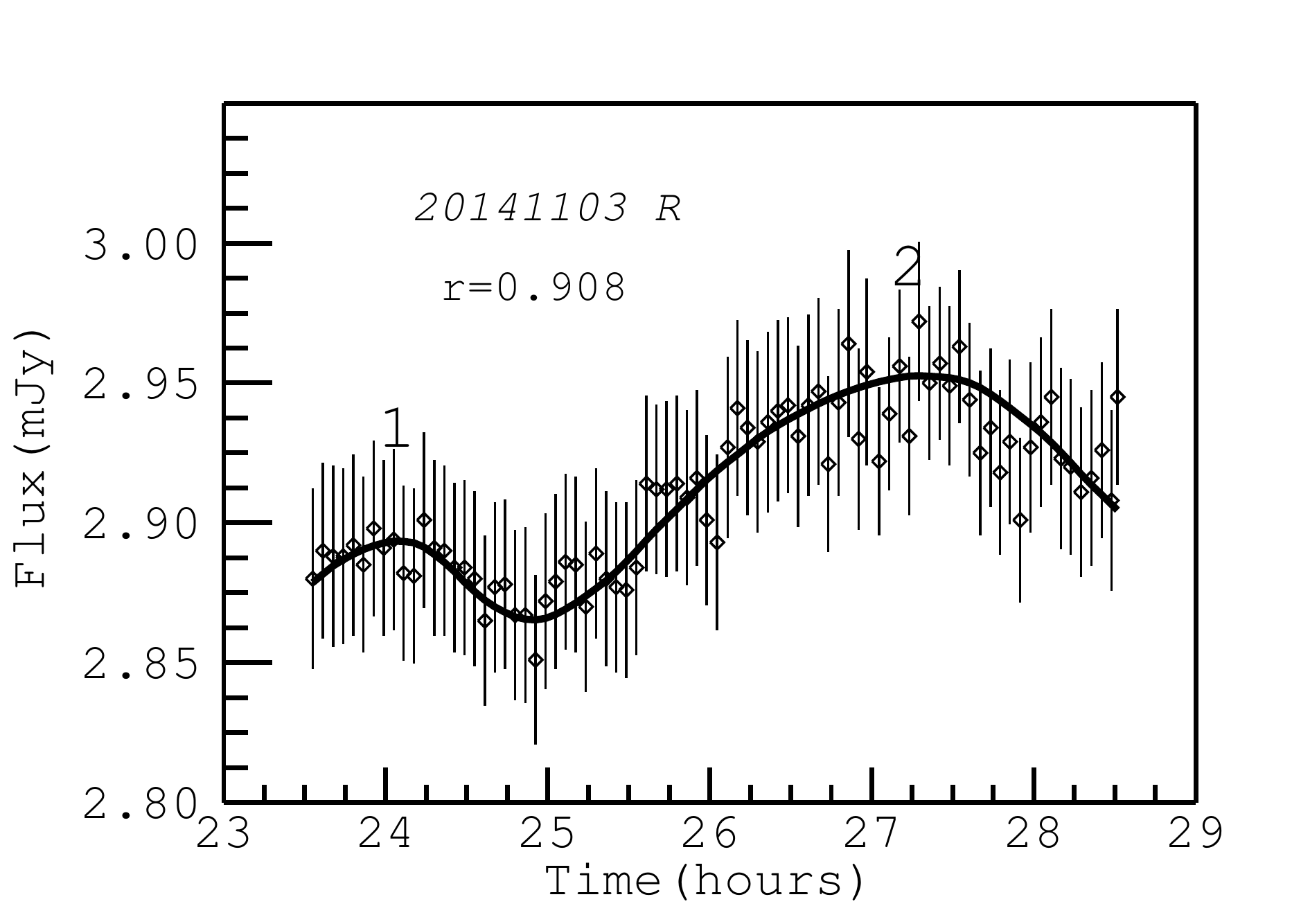}
	\includegraphics[scale=0.36]{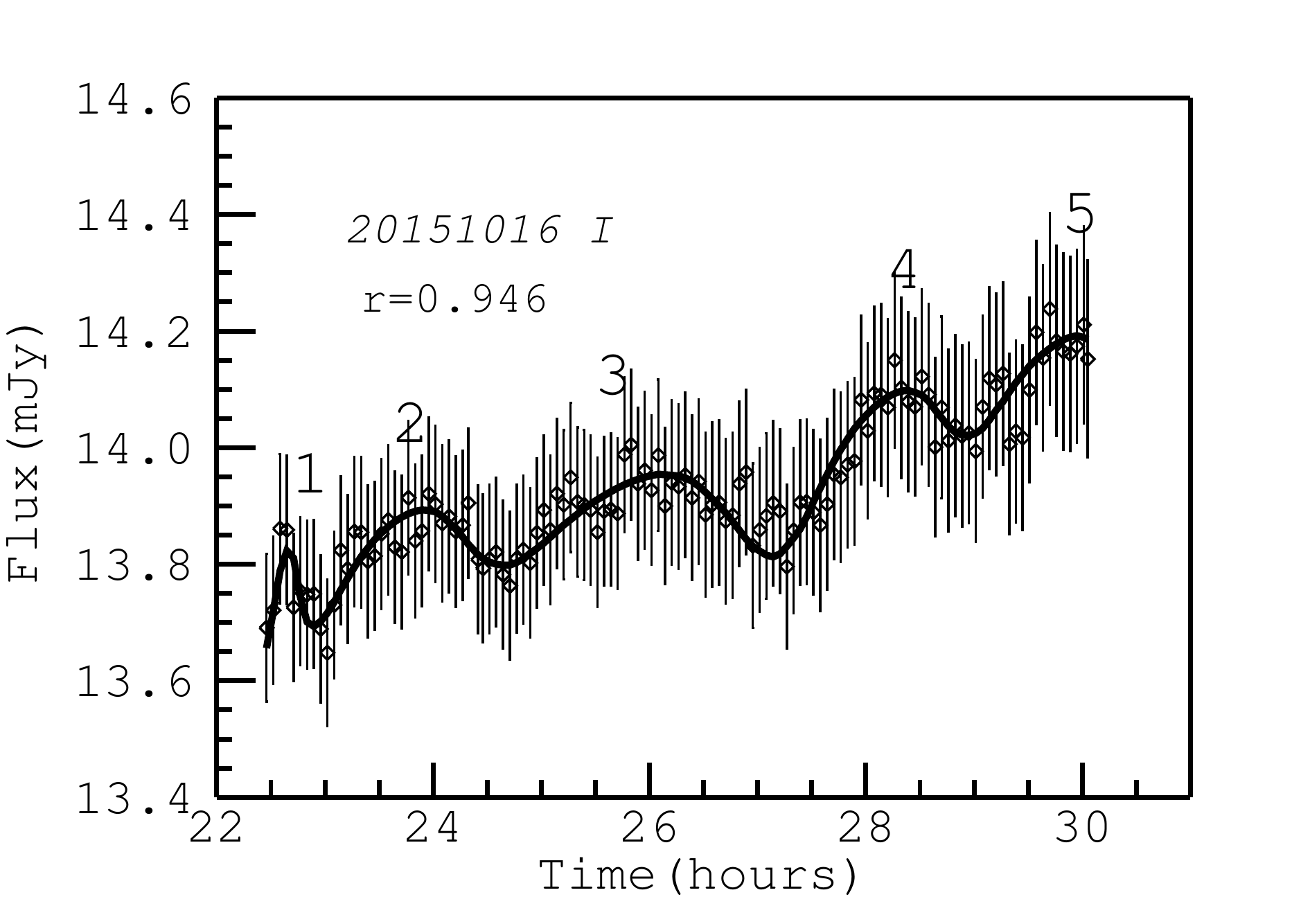}
	\includegraphics[scale=0.36]{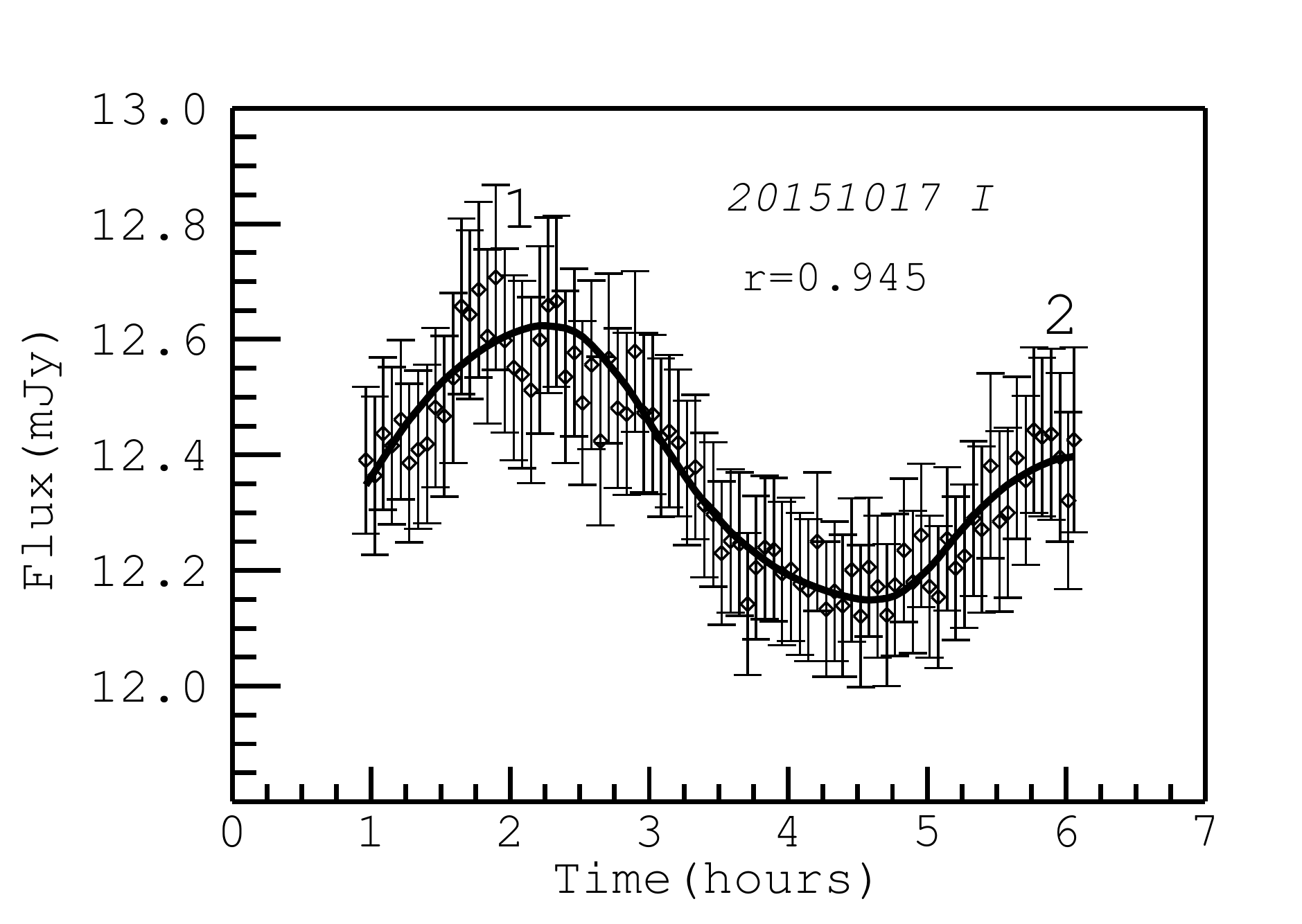}

      \caption{Light curves fitted with the convolutions of the synchrotron pulses of different amplitudes and widths listed in Table~\ref{tab:fits}.}
      \label{fig:pulse}
\end{figure}

\begin{table*}
	\centering
	\caption{Pulse parameters used to fit the data.}
	\label{tab:fits} 
	
	\begin{tabular}{cccrccr} 
		\hline
	         Date  &Band & Pulse & Centre & Amp  & $\tau_{\rm pulse}$    & $S_{\rm cell}$  \\
	         ( yyyy mm dd ) &  &    &  ( h ) & ( mJy ) & ( h ) & ( AU ) \\
		\hline
		2013 10 20 & $V$    &  1    &   22.80    &  0.19      &  0.84          &    4.40\\
		                   &       &   2   &   24.50    &   0.24     &   1.17         &    6.16\\
		                   &       &   3   &   25.50    &   1.54   &   1.03         &    5.43 \\
		                   &       &   4   &   27.38    &    0.24    &   1.39         &   7.34 \\
		                   &       &   5   &   28.41    &    0.18    &   0.56         &   2.93 \\
		                   &  $R$   &  1   &   22.64    &   0.23     &   1.11         &    5.87 \\
		                   &       &   2   &   24.39    &   0.39     &   1.20        &    6.31\\
		                   &       &   3   &   25.39    &   1.63     &   1.06         &    5.57\\
		                   &       &   4   &   27.25    &   0.38     &   1.34         &    7.04\\
	        2014 11 03 & $R$    & 1    &    24.00    & 0.05       &   1.36        &    7.19\\
	                           &       &   2   &   27.12    &   0.12      &   3.34         &    17.60\\
	        2015 10 16 &$ I$     & 1     &   22.65    &   0.20     & 0.28            &     1.47 \\
	                           &       &   2   &   23.82    &   0.28     &   1.53         &    8.07 \\
		                   &       &   3   &   26.00    &   0.35    &   2.48         &    13.06\\
		                   &       &   4   &   28.30    &   0.50    &   1.75         &   9.24\\
		                   &       &   5   &   29.90    &   0.60    &   1.67         &   8.80\\
	        2015 10 17 &$ I$      & 1    &  2.10         &  0.58       &  2.62       &  13.79\\
	                           &       &   2   &   6.00        &   0.32     &   1.87         &   9.83\\
		\hline
	\end{tabular}
\end{table*}

\subsection{Pulse Analysis}

Combining the results of the  $\chi ^2$ test with those of the ANOVA test, we extracted five IDV light curves with several obvious flares to test the theoretical model, which was investigated by \citet{2013A&A...558A..92B}. 
Based on the light-curve profile given by \citet{1998A&A...333..452K}, they assumed a turbulent jet to explain the microvariability of blazar S5 0716+714. 

As the strong shock hits each stochastic cell, particle acceleration and subsequent cooling by synchrotron emission produce a pulse. The convolution of these individual pulse emissions from inhomogeneous cells of various sizes and density enhancements leads to the observed microvariability.

For every local peak in the light curve, its location was taken as the centre position of the cell, its amplitude as the degree of the density enhancement and its width as its spatial extension. 
We applied this model to five intraday light curves by using their pulse code. We used the Doppler factor of 7.3 to calculate the pulse shape \citep{2009A&A...494..527H}. 

By varying the width and amplitude of the standard pulse, we have fitted each significant flare of the light curves. The resulting parameters for the pulses used in modelling the light curves are listed in Table~~\ref{tab:fits}. The first column shows the observation date, while the second column records the band. The pulse IDs and the centre time of the pulse are given in the 3rd and 4rd columns respectively, and the amplitude is in the 5th column. Column 6 gives width ( $\tau_{\rm pulse}$ ) of each pulse. The number quoted in Column 7 is an estimate of the size of the cell in AU based on the assumed shock speed ($u_ {\rm s}=0.1c$) and the duration of the pulse. Fig.~\ref{fig:pulse} shows the light curves fitted with the convolved pulses. Although the fit is not unique, it is representative of how well the model compares to the data. The correlation coefficient ($r$) of each fitting is calculated. On 2013 October 20, the peaks were missing in both bands; therefore, the correlation coefficients of fitting are less than 0.9. Based on the centre time of each pulse on 2013 October 20, the time lag is estimated to be about 8 min, which is consistent with the FR/RSS result. The other 3 d are well fitted with the model.

\section{Discussion}

There are various models to explain the IDV flux of blazars. Intrinsic ones include the instabilities in accretion disc \citep{1996ASPC..110...42W} and the shocks travelling down the jet  \citep[e.g.][ and references therein]{1996ASPC..110..248M,2014ApJ...780...87M}. Extrinsic ones involve gravitational microlensing \citep{1987A&A...171...49S} and interstellar scintillation \citep{2003ApJ...585..653B}.

Following the turbulent jet model, \citet{2013A&A...558A..92B} interpreted the microvariability as emission from individual synchrotron cells, which are energized by a plane shock propagating down the jet. This results in an increase in flux resembling a pulse. Since turbulence is a stochastic process, each microvariability curve is a realization of it. We use five IDV light curves with several flares to test their theoretical model and get the turbulent parameters from our observations. There is a large range of length scales for the turbulent vortices. 
The largest cell size is $\sim$17.6 AU that could correspond either to the correlation length scale or to the physical width of the jet, while the smallest cell size is around 1.5 AU, could correspond to the Kolmogorov scalelength of the turbulent plasma. We can get a picture of the underlying turbulent structure.
Fitting pulses to BL Lacertae microvariability curves give us a much better indication of the turbulent nature of the plasma in these sources.

\citet{1989Natur.337..627M} argue that the microvariations are produced very close to the central supermassive black hole (BH). Several works have attempted to estimate the mass of the BH in BL Lacertae   \citep{1999A&AS..136...13F,2002ApJ...579..530W,2010MNRAS.402..497G,2010A&A...516A..59C,2012NewA...17....8G}, which seems to be $0.1-6$ $\times 10^8  \rm M_{\odot}$. If we assume fluctuations in the inner portions of the accretion disc, the observed minimum time-scale $\Delta t_{\rm obs}$ will provide an upper limit to the mass of BH.
\citet{2010MNRAS.404..931E} pointed out that the first-order structure function (SF) sometimes leads to incorrect claims of time-scales. Hence, we adopt the ZDCF (in autocorrelation mode) method to get a possible IDV time-scale. We choose the minimum zero-crossing time of the DCF as the correlation time-scale and get the time-scale of variability of 42.5 mins on 2013 October 20 (in Fig.~\ref{fig:BHmass}). 
Then according to \citet{2012NewA...17....8G}, the mass of BH can be estimated by,
\begin{equation}
   M_{\rm BH}=\frac{c^3\Delta t_{\rm obs}}{10G(1+z)} ,
\end{equation}
For our target, $ M_{\rm BH}$ is calculated to be 0.49$\times 10^8  \rm M_{\odot}$. If the variations arise in the jets and are not explicitly related to the inner region of the accretion disc, the BH mass estimation is invalid. 

We might try to launch possible multiwavelengths observation campaign in the future to gain a much more comprehensive understanding of the physical model of blazars.

\section{Conclusions}

Our conclusions are summarized as follows:

\begin{itemize}
\item We carried out a four-colour monitoring programme on BL Lacertae from 2012 to 2016. The variations were well correlated in all bands.
\item The amplitude of IDV is greater in higher energy bands.
\item After the host galaxy contribution was removed, the source exhibits a BWB trend. The spectral indices varied slightly. 
\item The possible time delays are about 10 min between variations in the $V$ and $R$ bands. The IDV light curves with flares are helpful to study time-lag detection. Further observations programme with a temporal resolution of about 1 min are needed to validate our time-lag results. 
\item Our data can be well fit by the model of individual synchrotron pulses.

\end{itemize}

\section*{Acknowledgements}

We would like to thank the anonymous referee for insightful comments and suggestions to improve this manuscript. 
This work has been supported by the National Basic Research Programme of China 973 Program 2013CB834900; Chinese National Natural Science Foundation grants 11273006 and U1531242; and the fundamental research funds for the central universities and Beijing Normal University.




\bibliographystyle{mnras}








\bsp	
\label{lastpage}
\end{document}